\documentclass[12pt]{article}
\usepackage[margin=1.0in]{geometry}
\usepackage[printwatermark]{xwatermark}

\usepackage{verbatim}
\usepackage{amsmath}
\usepackage{multirow}
\usepackage{apacite}

\usepackage{xcolor}
\usepackage{tabu}
\usepackage{float}
\usepackage{graphicx}

\usepackage[utf8]{inputenc}
\usepackage{gensymb}
\usepackage[section]{placeins}
\usepackage{natbib}
\usepackage{lineno}
\usepackage{authblk}

\usepackage{titlesec}
\titleformat*{\section}{\large\bfseries}
\titleformat*{\subsection}{\normalsize\bfseries}

\usepackage{setspace}
\usepackage{etoolbox}
\preto\longtable{\par\singlespacing}

\usepackage{hyperref}
\hypersetup{
	colorlinks = true,
	linkcolor = {red},
	citecolor = {blue},
	linkbordercolor = {white},
	citebordercolor = {purple}
}

\title{Ejecta distribution and momentum transfer from oblique impacts on asteroid surfaces}

\author[1, 2]{S. D. Raducan}
\author[1]{T. M. Davison}
\author[1]{G. S. Collins}
\affil[1]{Impacts and Astromaterials Research Centre, Department of Earth Science and Engineering, Imperial College London, United Kingdom.}
\affil[2]{Space Research and Planetary Sciences, Physikalisches Institut, University of Bern, Switzerland;}

\date{}
\begin{document}

\maketitle

\maketitle
\begin{abstract}
NASA's Double Asteroid Redirection Test (DART) mission will impact its target asteroid, Dimorphos, at an oblique angle that will not be known prior to the impact. We computed iSALE-3D simulations of DART-like impacts on asteroid surfaces at different impact angles and found that the the vertical momentum transfer efficiency, $\beta$, is similar for different impact angles, however, the imparted momentum is reduced as the impact angle decreases. It is expected that the momentum imparted from a 45$^\circ$ impact is reduced by up to 50\% compared to a vertical impact. The direction of the ejected momentum is not normal to the surface, however it is observed to `straighten up' with crater growth. 
iSALE-2D simulations of vertical impacts provide context for the iSALE-3D simulation results and show that the ejection angle varies with both target properties and with crater growth. While the ejection angle is relatively insensitive to the target porosity, it varies by up to 30$^\circ$ with target coefficient of internal friction. The simulation results presented in this paper can help constrain target properties from the DART crater ejecta cone, which will be imaged by the LICIACube. 
The results presented here represent the basis for an empirical scaling relationship for oblique impacts and can be used as a framework to determine an analytical approximation of the vertical component of the ejecta momentum, $\beta-1$, given known target properties.
\end{abstract}

\section{Introduction}

NASA's Double Asteroid Redirection Test (DART) will be the first mission to test a controlled deflection of a Near-Earth binary asteroid \citep{Cheng2016, Michel2016}, by impacting the smaller component of the 65803 Didymos asteroid system, Dimorphos. The impact will thereby alter the binary orbit period by an amount detectable from Earth \citep{Cheng2018}. ESA's Hera mission \citep{Cheng2018, Michel2018} will arrive at Dimorphos several years after the DART impact, rendezvous with the asteroid system and perform detailed characterisation of Dimorphos and the DART impact outcome. Hera will determine the target's volume and surface properties, as well as the volume and morphology of the DART impact crater and the post-impact dynamical state of the Didymos system.

In a high-velocity, head-on impact between a spacecraft and an asteroid, the change in momentum of the asteroid, $\Delta P$, can be amplified by the momentum of crater ejecta that exceeds the asteroid’s escape velocity. The total momentum change to the asteroid divided by the impactor momentum is a measure of deflection efficiency, commonly defined as $\beta = \Delta P/(mU)$,  where $mU$ is the impactor momentum. $\beta = 1$ implies that the crater ejecta makes a negligible contribution to the total momentum change, while $\beta>2$ means that the momentum contribution from the crater ejecta is larger than the momentum imparted by the impactor directly. The amount by which crater ejecta enhances asteroid deflection---that is, the momentum of the crater ejecta that escapes the gravitational attraction of the target body divided by the impactor momentum ($\beta-1$)---has been found to vary significantly depending on the target asteroid’s properties, composition and structure \citep[e.g.,][]{Jutzi2014, Stickle2015, Syal2016, Raducan2019, Raducan2020}. 
These numerical studies have considered DART as a vertical impact, however, in reality, the DART spacecraft will likely impact Dimorphos' surface at an oblique angle \citep{Cheng2018}. The exact impact angle will depend on both the spacecraft's incoming trajectory \citep{Atchison2016} and on the local slope of the target at the impact point. While the spacecraft's trajectory can be computed to some degree of accuracy, the target surface morphology is not known prior to the impact. 

Even though it is not yet well understood how $\beta-1$ is affected by the impact angle, it is expected that any departure from a vertical impact will reduce the deflection efficiency. This trend was seen previously in laboratory experiments  \citep[e.g.,][]{Yanagisawa2000}, however  the effects of impact angle on momentum transfer are yet to be fully quantified. Several laboratory experiments have measured the ejection angle and ejection velocities of ejecta produced by oblique impacts \citep{Anderson2003, Anderson2004}, however such studies are difficult to conduct in laboratory and only limited data is available. The effect of impact angle on ejecta mass-velocity-launch position distributions, and its interplay with target properties, has not been systematically investigated, as this requires much more computationally demanding fully three-dimensional calculations. Numerical simulations can be used to quantify the effects of various target properties and various impact conditions on the cratering process \citep[e.g.,][]{Prieur2017, Luther2018, Raducan2019} and inform empirical scaling laws. For example, \cite{Raducan2019} showed that empirical scaling relationships for vertical impacts \citep{Housen2011} can be used to accurately predict momentum transfer, however equivalent scaling relationships for oblique impacts do not exists.

In this work we simulated the DART experiment as an oblique impact at four different impact angles in three dimensions. We use our simulation results to determine the net momentum transfer of the DART impact for one possible set of asteroid target properties. Our simulation results are also used to determine the radial and azimuthal variation in ejection velocity and angle for different impact angles. Together with vertical simulation results of the radial variation in ejection velocity and angle, as a function of different target properties, these data allow us to  develop a framework for integrating the effect of impact angle into existing crater ejecta scaling relationships and for estimating the deflection efficiency of an oblique impact, given known target properties and impact conditions.

\section{Scaling of crater size and ejecta mass-velocity distribution from vertical impacts}

Crater and ejecta scaling relationships are widely used to extrapolate the results of impact experiments at laboratory-scale to predict the outcome of planetary-scale events. The most widely used of these relationships are derived by approximating the impact as a point source \citep{Holsapple1987}. In this case, any outcome of an impact, such as crater size or mass-velocity distribution of the crater ejecta, is related to impactor and target properties (e.g., impactor mass, speed, target cohesion, porosity) through a so-called coupling parameter $C\sim a\delta^\nu U^\mu$  that represents the impactor's influence on the cratering process in a single measure \citep{Housen1983, Housen2011}. In this expression, $a$ is the impactor radius, $\delta$ is the impactor density, $U$ is the impactor speed and $\nu$ and $\mu$ are constants. The density scaling exponent $\nu$ is often assumed to be independent of material properties, with a value of $\approx$ 0.4 \citep{Schmidt1980, Housen2003}. On the other hand, the velocity scaling exponent $\mu$ depends on the target material properties \citep{Schmidt1987, Prieur2017, Raducan2019} and takes values between the theoretical limits of $\mu$ = 1/3 if the coupling parameter scales with the momentum of the impactor, and $\mu$ = 2/3 if the coupling parameter scales with the impact energy \citep{Holsapple1987}. 

For example, applying the coupling parameter concept to crater size, the crater radius, $R$, normalised by the cube root of the target density, $\rho$, and impactor mass, $m$, can be expressed in terms of the $\pi$-group dimensionless parameters $\pi_2 = \frac{ga}{U^2}$, $\pi_3 = \frac{Y}{\rho U^2}$ and $\pi_4 = \frac{\rho}{\delta}$, where $g$ is the target gravity and $Y$ is a measure of target strength. If crater growth is halted principally by the target strength, then crater formation is said to occur in the `strength' regime. In this case, crater radius is independent of $\pi_2$ and the scaling relationship takes the form: 
\begin{linenomath*}
\begin{equation}\label{eq:pi_2}
R\left(\frac{\rho}{m}\right)^{1/3} = H_{2} \left(\frac{Y}{\rho U^2}\right)^{-\mu/2} \left(\frac{\rho}{\delta}\right)^{(1-3\nu)/3} \ \ \ \mathrm{(strength).}
\end{equation}
\end{linenomath*}
where $H_2$ is a scaling constant. For large craters or weak materials, where crater growth is controlled principally by the target gravity, crater radius is instead independent of $\pi_3$ and the scaling relationship takes the alternative form:
\begin{linenomath*}
\begin{equation}\label{eq:pi_3}
R\left(\frac{\rho}{m}\right)^{1/3} = H_{1} \left(\frac{ga}{U^2}\right)^{-\frac{\mu}{2+\mu}}\left( \frac{\rho}{\delta} \right)^{\frac{2+\mu-6\nu}{3(2+\mu)}} \ \ \ \mathrm{(gravity).}
\end{equation}
\end{linenomath*}
where $H_1$ is a scaling constant. 

Eqs.~\eqref{eq:pi_2}--\eqref{eq:pi_3} apply to vertical impacts only.  
Previous numerical studies and laboratory experiments of oblique impacts suggest that crater volume and crater diameter decrease with decreasing impact angle, in a manner that is approximately consistent with the idea that only the vertical component of the impact velocity ($U \sin(\theta)$) contributes to the growth of the crater in an oblique impact \citep{Chapman1986, Elbeshausen2009, Davison2011}.  

Using the same coupling parameter concept and point-source approximation, \cite{Housen2011} developed a number of power-law scaling equations that relate properties of ejecta to the initial vertical impact conditions. In one such relationship, the speed of ejecta, $v$, is expressed as a function of ejecta launch position, $r$, as well as impactor and target properties \citep{Housen1983, Housen2011}:

\begin{linenomath*}
\begin{equation}\label{eq:v_x}
\frac{v(r)}{U}=C_1\left[\frac{r}{a}\left(\frac{\rho}{\delta}\right)^\nu\right]^{-\frac{1}{\mu}}\left(1-\frac{r}{n_2R}\right)^p,
\end{equation}
\end{linenomath*}
where $C_1$ and $p$ are material dependant fitting constants. The relationship is also invalid for very fast ejecta, where $r < n_1a$ and $n_1 \approx 1.2$. To extend the relationship closer to the impact point, \cite{Raducan2019} proposed an additional empirical term, $\left(1-\frac{r}{n_1a}\right)^q$, that includes the fast ejecta behaviour:
\begin{linenomath*}
\begin{equation}\label{eq:v_x2}
\frac{v(r)}{U}=C_1\left[\frac{r}{a}\left(\frac{\rho}{\delta}\right)^\nu\right]^{-\frac{1}{\mu}}\left(1-\frac{r}{n_2R}\right)^p\left(1-\frac{r}{n_1a}\right)^q,
\end{equation}
\end{linenomath*}
where $q$ is a target and material dependent constant. \cite{Raducan2019} found that for spherical aluminium projectiles at moderate velocities ($U\approx7$ km/s), the constant $q$ takes values close to 0.2. However, further studies are needed to determine the influence of target and projectile properties on $q$.

The ejection speed decreases as the launch distance $r$ increases. \citet{Housen2011} defined the mass ejected at speeds larger than $v$, $M(>v)$, as the mass of material $M(<r)$ launched at distances from within $r$:
\begin{linenomath*}
\begin{equation}\label{eq:m_x}
\frac{M(<r)}{m} = \frac{3k}{4\pi}\frac{\rho}{\delta}\left[\left(\frac{r}{a}\right)^3-n_1^3\right].
\end{equation}
\end{linenomath*}
where $k$ is a material fitting constant. 

As with crater size scaling, no definitive extension to these ejecta scaling relationships exists for oblique impacts. Here we present some preliminary steps towards such an extension, based on our oblique impact simulation results.

\section{Numerical model}

To model the DART impact into possible asteroid surfaces we used the iSALE shock physics code \citep{Wunnemann2006} in two and three dimensions. The iSALE shock physics code is a multi-material, multi-rheology extension of the SALE (Simplified Arbitrary Lagrangian Eulerian) hydrocode \citep{Amsden1980} that was specifically designed for simulating impact processes and is similar to the older SALEB hydrocode \citep{Ivanov_Artemieva2002, Ivanov1997}. iSALE-3D \citep{Elbeshausen2009, Elbeshausen2011} uses a 3D solution algorithm very similar to the SALE-2D solver, as described by \cite{Hirt1974}. The development history of iSALE-3D is described in \cite{Elbeshausen2009}. Both codes share the same material modelling routines, including strength models suitable for impacts into geologic targets \citep{Collins2004} and a porosity compaction model \citep{Wunnemann2006}. 
The crater sizes produced by iSALE-3D simulations of oblique impacts into aluminium targets agree well with laboratory experiments \citep{Davison2011}, while the ejection velocities and angles of ejecta produced by vertical impacts are in good agreement with data from laboratory impacts into sand \citep{Luther2018, Raducan2019}.

Here we aim to quantify the effect of impact angle on impact momentum transfer and ejecta mass and velocity distributions. Since full 3D simulations are very computationally expensive we performed a relatively small number of oblique impact simulations into targets with constant properties, for which we varied the impact angle. To get additional insight into the effects of key material properties on ejection angles we also revisited the \citet{Raducan2019} 2D simulations results. \citet{Raducan2019} studied the effects of material properties (cohesion, porosity and coefficient of internal friction) on ejection velocity and ejected mass in analog DART impacts. Here we re-analyse these results to quantify the dependence of ejection angle as a function of radius on target properties. 

All the numerical simulations presented here used a $\approx$620 kg spherical aluminium projectile, impacting a porous, basaltic regolith target at 7 km/s. As in \cite{Raducan2019}, the impactor was modelled using the Tillotson equation of state for aluminium \citep{Tillotson1962} and a simple von Mises strength model. 

\begin{table}[h]
    \footnotesize
	\caption{Material model parameters for impact simulations into Dimorphos analogues. For all simulated materials we used the thermal parameters from \cite{Ivanov2010}. }
	\begin{tabular}{l@{\hskip 0.1in}l@{\hskip 0.2in}l@{\hskip 0.2in}l}
    Description        & Impactor & iSALE-2D target & iSALE-3D target  \\
    \hline
    Material             & Aluminium & Basalt & Basalt\\
    Impact angle ($\circ$)  & -- & 90     & 90/60/45/30 \\
    Impact speed (km/s)  & -- & 7      & 7 \\
	\hline
	Equation of state        & Tillotson$^{a}$ & Tillotson$^b$  & Tillotson$^b$\\
	Strength model           & von Mises & LUND$^{c, d}$ & LUND$^{c, d}$  \\
	Poisson ratio$^c$, $\nu$ & 0.33      & 0.25     & 0.25 \\
	\hline
	Thermal parameters   \\     
	Melting temperature, T$_m$ (K)  &  933    & 1360   & 1360 \\
	T$_{frac}$                      &  1.2    &  0.7   & 0.7  \\
	A$_{simon}$ (GPa)               &  6.0    &  4.5   & 4.5  \\
	C$_{simon}$                     &  5.00   &  2.11  & 2.11 \\ 
	\hline
	LUND strength parameters$^{c, d}$ \\
	Damage strength at zero pressure, Y$_0$ (kPa)   & --  & 10  & 10 \\
	Strength at infinite pressure, Y$_{\inf}$ (GPa) & --  & 1  & 1 \\
	Internal friction coefficient (damaged), $f$    & --  & 0.2--1.2  & 0.6 \\
	\hline
	Porosity model parameters ($\epsilon-\alpha$)$^e$           \\ 
	Initial porosity, $\phi_0$             & 60\% & 10--50\% & 20\% \\
	Initial distension, $\alpha_0$         & 2.7 & 1.1--2.0 & 1.67 \\
	Distension at transition to power-law, $\alpha_x$ & 1.00 &  1.00 & 1.00 \\
	Elastic volumetric strain threshold, $\epsilon_{e0}$ & 0.00 & -1.88$\times 10^{-6}$ & -1.88$\times 10^{-6}$ \\
	Exponential compaction rate, $\kappa$        & 0.9 & 0.88--0.98 & 0.90 \\
	\hline
    \multicolumn{4}{l}{
    $^a$\cite{Tillotson1962};
    $^b$\cite{Benz1999};
    $^c$\cite{Lundborg1967};
    $^d$\cite{Collins2004};}\\{
    $^e$\cite{Wunnemann2006}.}\\
	\end{tabular}
	\label{table:model_parameters}
\end{table}
\FloatBarrier

The porous basaltic target model used here is considered to be a good approximation of the compositional structure of most asteroids. It comprises a Tillotson equation of state for basalt \citep{Tillotson1962, Benz1999} to describe the volumetric response of the solid of the target.  The porosity compaction behaviour of the target material was described using the $\epsilon-\alpha$ model \citep{Wunnemann2006, Collins2011}. The $\epsilon-\alpha$ input parameters were chosen so that the target crush-curve is consistent with shock wave and Hugoniot data for regolith-like materials \citep{Ahrens1974, Raducan2020}. The target's shear strength was modelled using a simple pressure-dependent strength model typical of geologic materials \citep{Lundborg1967, Collins2004}, with a damaged strength at zero pressure, $Y_0$ = 10 kPa. The impactor and target material properties are summarised in Table~\ref{table:model_parameters} and have been described in detail in \cite{Raducan2019}.

To record the impact ejecta we followed the same approach as in \cite{Raducan2019}: Lagrangian tracer particles were placed across the high-resolution domain and their mass, velocity vector and launch position were then recorded at a fixed altitude, which was set to one impactor diameter. In the 3D simulations, launch position and azimuth were measured relative to the impact point.

\subsection{iSALE-2D vs iSALE-3D}

To verify consistency between iSALE-2D and iSALE-3D for predictions of ejected mass and momentum (i.e., that the results are independent of model geometry) we used both codes to simulate the same vertical impact scenario. The scenario considered was a DART impact into a homogeneous half-space, with the cohesive strength of the damaged material, $Y_0$ = 10 kPa, coefficient of internal friction, $f$ = 0.6, and initial porosity, $\phi_0$ = 20\%. 

\begin{figure}[!h]
	\centering
	\includegraphics[width=\linewidth]{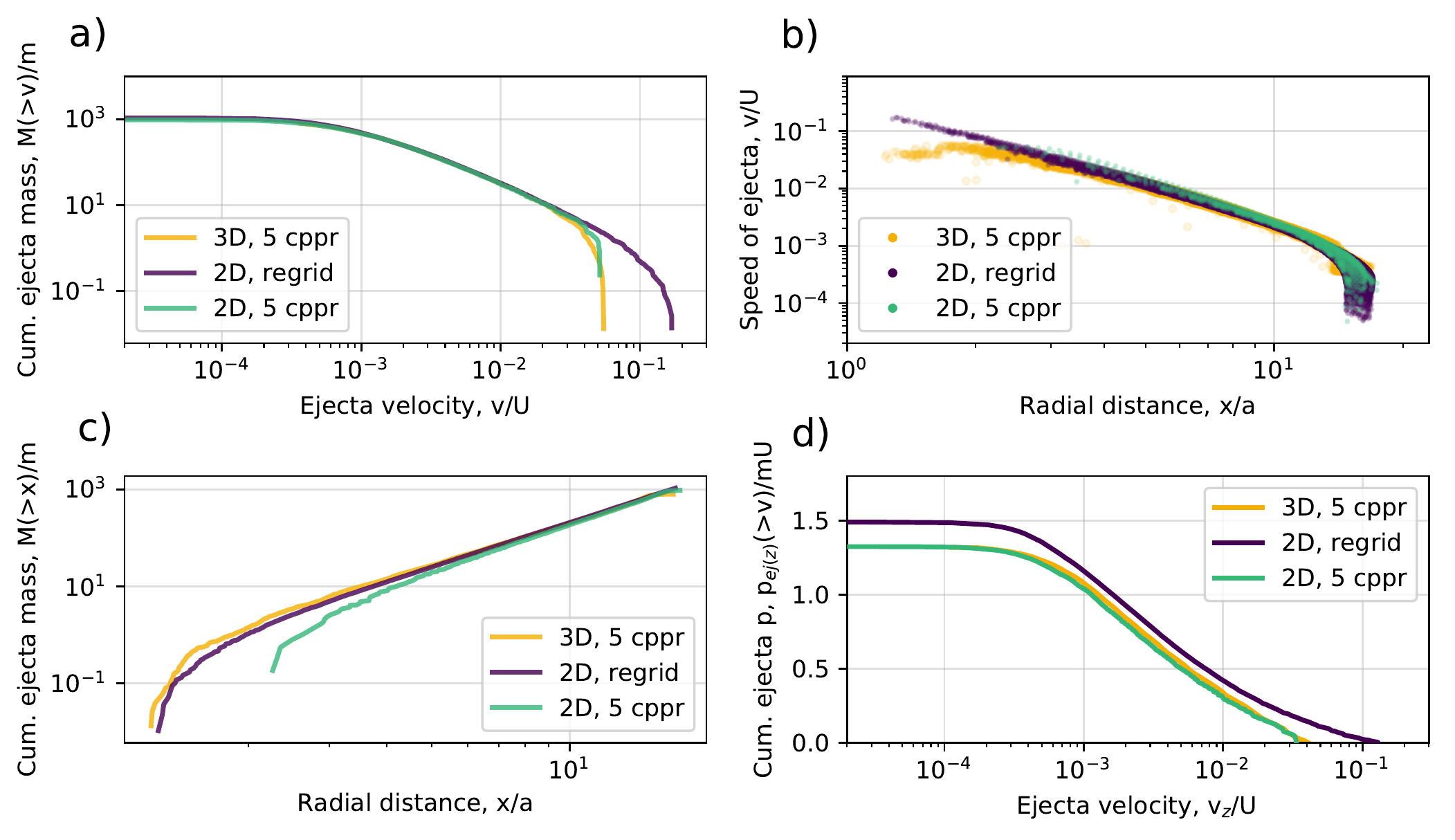}
	\caption{Mass-velocity-launch position distribution of ejecta from iSALE-2D simulations at 40 cppr and 5 cppr, compared with ejecta distribution from iSALE-3D at 5 cppr.} 
	\label{fig:ejecta_90}
\end{figure}

Figure~\ref{fig:ejecta_90} shows the mass-velocity-launch position ejecta distributions from three numerical simulations: two iSALE-2D runs and one iSALE-3D run. The three-dimensional (3D) simulation was limited to a spatial resolution of 5 cppr (3D, 5 cppr). For direct comparison, a 2D simulation of the same spatial resolution was performed (2D, 5 cppr). To assess the sensitivity of the results to the low spatial resolution of the 3D simulation, we also performed a 2D simulation that began with a resolution of 40 cppr and was subsequently coarsened using regridding to expedite the calculation without significant loss of accuracy (2D, regrid) as adopted in previous work \citep{Raducan2019, Raducan2020} and the other 2D simulations presented here. Based on previous resolution studies \citep{Raducan2019}, the lower spatial resolution used in the 3D simulations is expected to under predict the crater volume by about 5\% and the ejected momentum by about 10\% compared to the 2D simulations with regridding. 

Comparison between the 3D and 2D simulations with the same resolution (5 cppr) demonstrates consistency between iSALE-2D and iSALE-3D results (Fig.~\ref{fig:ejecta_90}). In particular, the cumulative ejecta mass-velocity distributions and the cumulative ejected momentum-velocity distributions are nearly identical for the full range of ejection velocities. When compared to the results of the high-resolution 2D simulation where regridding was used (2D, regrid), on the other hand, the 3D simulation results show good agreement in the cumulative mass and launch position of the slow ejecta, but fail to capture the fastest ejecta---known to require high spatial resolution at early times \citep{Johnson2014}. These fast particles, although of low mass, add $\approx$10\% to the normalised cumulative ejected momentum in this example (Fig.~\ref{fig:ejecta_90}d). As a result, the 3D simulation results presented here will systematically under-predict the cumulative ejected momentum.

\section{Results}

\subsection{Influence of impact angle on net momentum transfer}

Having verified the consistency between iSALE-2D and iSALE-3D simulation results for a vertical impact, we then used iSALE-3D to investigate the effect of impact angle on ejecta mass, velocity and angle distributions.

\begin{figure}[!h]
	\centering
	\includegraphics[width=\linewidth]{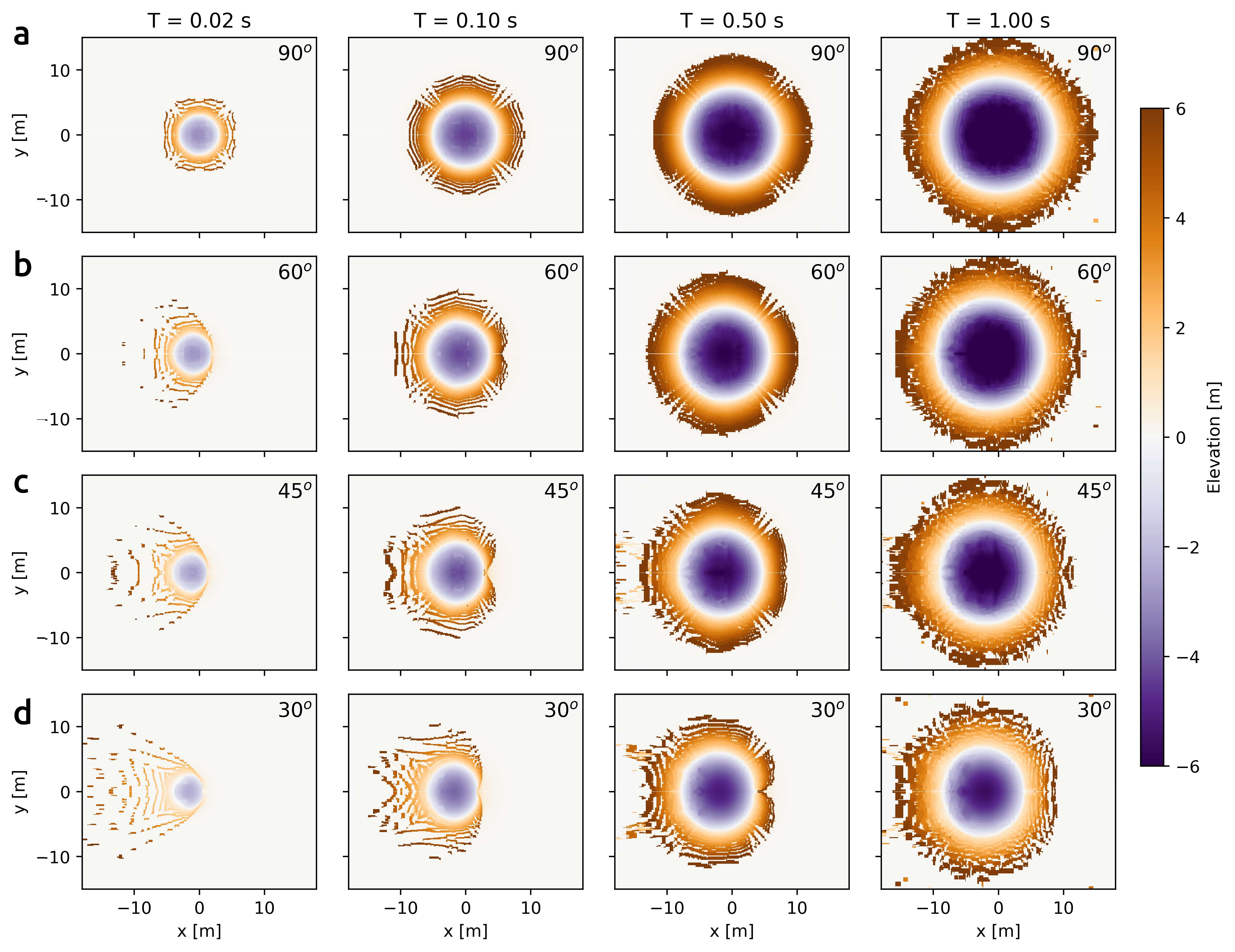}
	\caption{Surface topography of the DART impact at (a) 90, (b) 60, (c) 45 and (d) 30 degrees angles, captured at four different times: 0.02 s, 0.10 s, 0.40 s and 1.00 s. The impact direction is right to left. } 
	\label{fig:topography}
\end{figure}

Figure~\ref{fig:topography} shows the plan view of crater and ejecta curtain evolution following a DART impact at $90^\circ$ (vertical), $60^\circ$, $45^\circ$  and $30^\circ$ angles of incidence to the target surface. The impact speed is the same in all simulations ($U$ = 7 km/s); all other impactor and target properties are the same. The views are centered on the impact point (impactor tangent to the surface), with impact direction from right to left. The time frames of the oblique impacts (Figure \ref{fig:topography}b, c, d) show a highly asymmetric ejecta distribution at early times of the cratering process $<$ 0.10 s, compared to the same times in the vertical impact (Figure \ref{fig:topography}a). The asymmetric ejecta curtain becomes more symmetric as the crater grows towards its final size. 

\begin{figure}[!h]
	\centering
	\includegraphics[width=\linewidth]{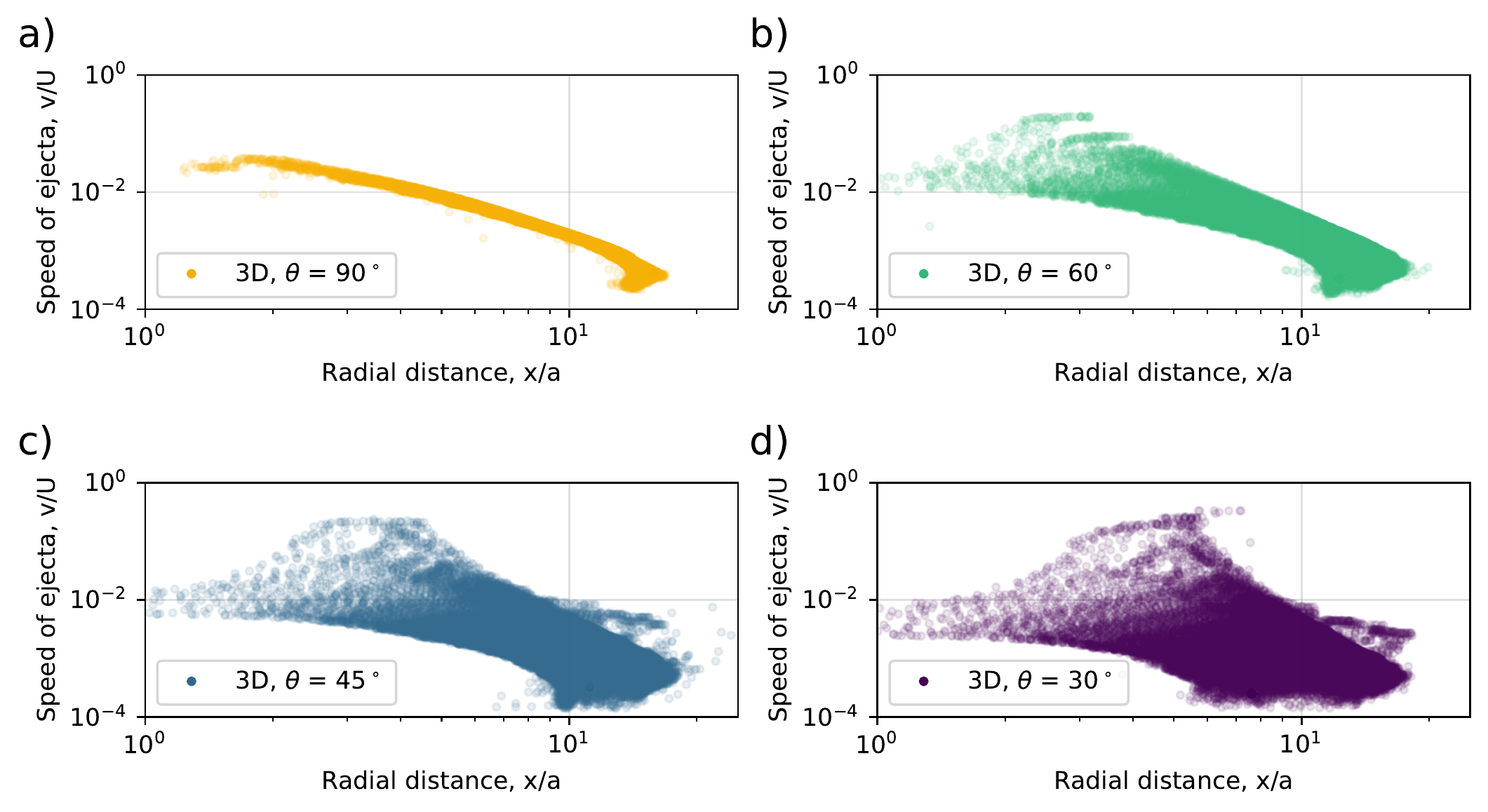}
	\caption{Velocity launch position distribution of the ejecta from impacts at 90, 60, 45 and 30 degrees angle of incidence.} 
	\label{fig:ejecta_all}
\end{figure}

The asymmetry in the ejecta is also illustrated by the velocity-launch position distribution of the ejected particles from the oblique impacts at 60, 45 and 30 degrees angle of incidence (Fig.~\ref{fig:ejecta_all}). In all cases, the launch speed of the ejecta, $v$, was normalised by the impact speed, $U$, and plotted as a function of launch position, $r$ (relative to the impact point), normalised by the impactor radius, $a$. The ejecta from the oblique impacts displays higher speeds in the downrange direction, and lower speeds in the uprange direction. This is consistent with previous laboratory-scale oblique impact experiments \citep{Schultz1999, Anderson2003} and DART impact models \citep{Stickle2015}.

\begin{figure}[h!]
	\centering
	\includegraphics[width=0.60\linewidth]{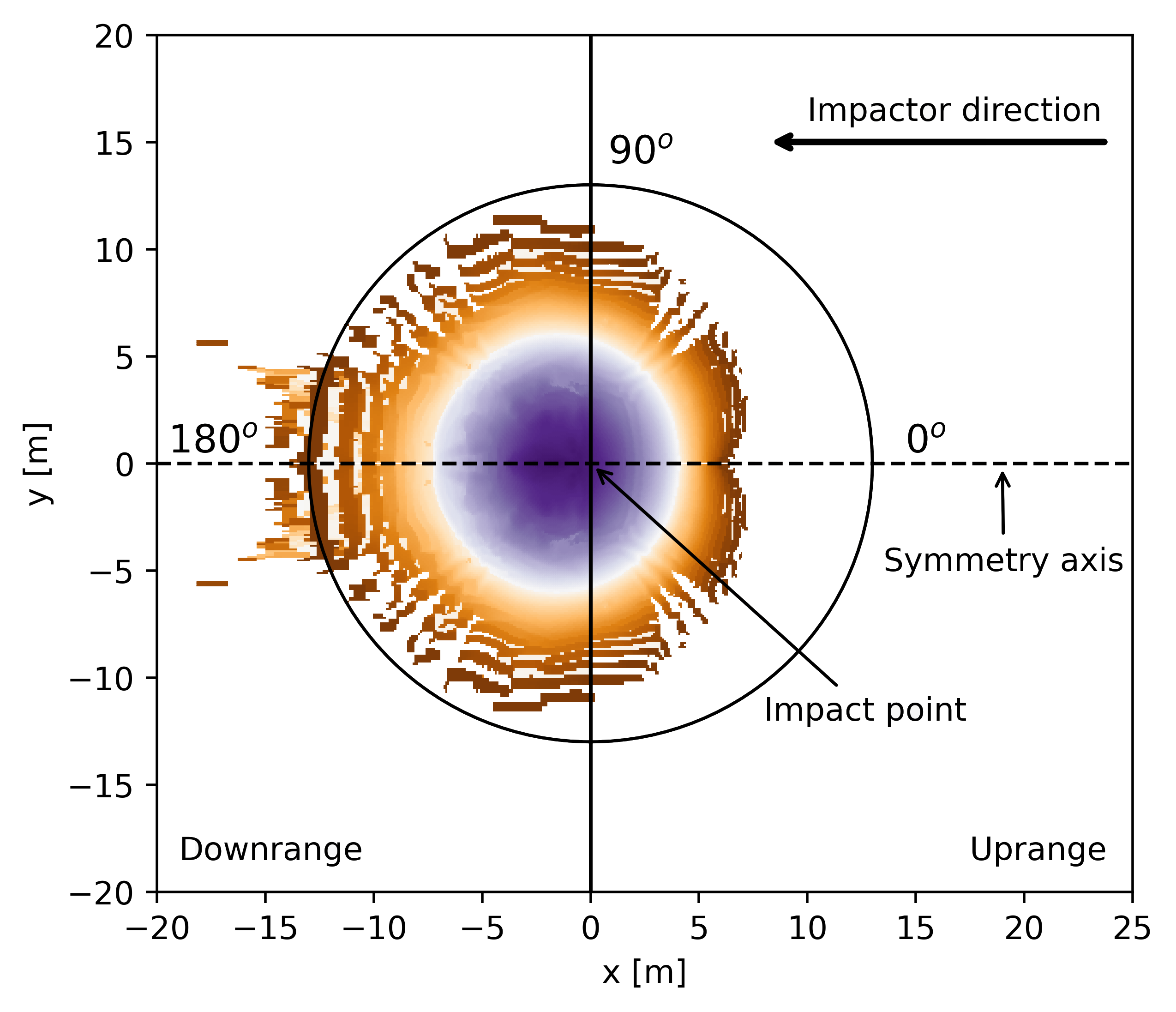}
	\caption{Surface topography of the DART impact at 45$^\circ$ angle of incidence, showing the direction of the impactor and the azimuthal ($\zeta$) coordinates relative to the impact point.}
	\label{fig:angle_map}
\end{figure}

Following a similar approach to \citet{Anderson2004} to better understand how the ejecta velocity and ejection angle vary with azimuth around the impact point, we split the ejecta velocity distribution into azimuthal sections, between between $\zeta = 0^\circ$, which represents the uprange direction ($-x$) and $\zeta = 180^\circ$, which represents the downrange direction ($+x$). Figure~\ref{fig:angle_map} shows the surface topography of the DART impact at 45$^\circ$ impact angle and a diagram of the locations of the impact point, symmetry axis and the azimuth angles. 

Unlike in the vertical impacts, in oblique impacts the centre of the crater is not stationary, but instead moves from the impact point towards the downrange direction as the crater grows. When determining the normalised radial launch position of the ejecta, $r/a$, the origin was defined as the impact point, rather than the centre of the final crater. When the data from all azimuths is aggregated, this convention causes a larger spread in launch distance for a given ejection velocity compared to the 90$^\circ$ impact, and must be accounted for in the ejecta analysis. It also implies that the outermost launch position, which defines the approximate edge of the crater, varies with azimuth, even if the crater has a circular planform.

\begin{figure}[!h]
	\centering
	\includegraphics[width=\linewidth]{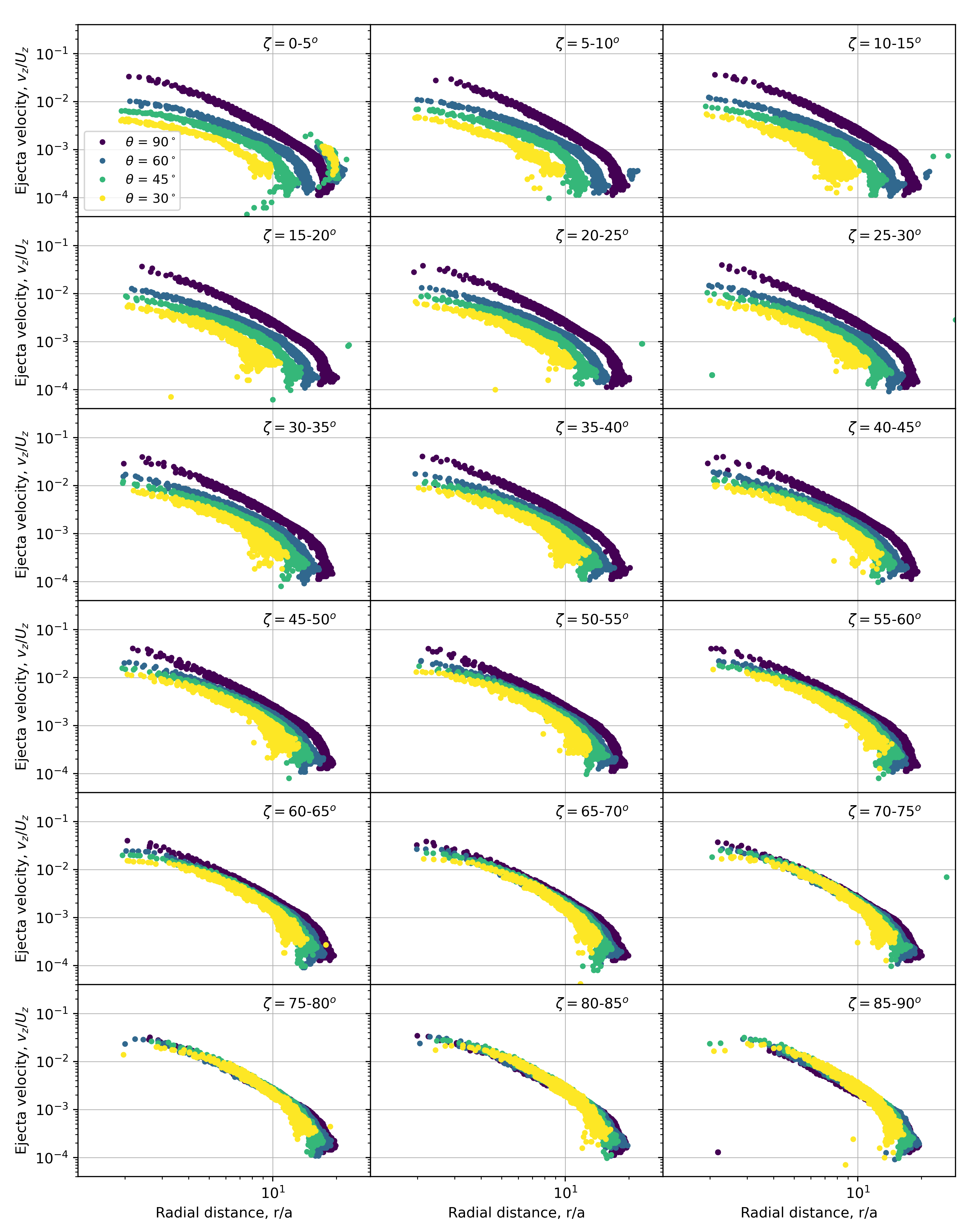}
	\caption{Ejecta vertical velocity - launch position distribution from vertical and oblique impacts ($\theta$ = 90$^\circ$, 60$^\circ$, 45$^\circ$, 30$^\circ$) at azimuth between $\zeta$ = 0 and $\zeta$ = 90$^\circ$.}
	\label{fig:azimuth_fit}
\end{figure}

\begin{figure}[!h]
	\centering
	\includegraphics[width=\linewidth]{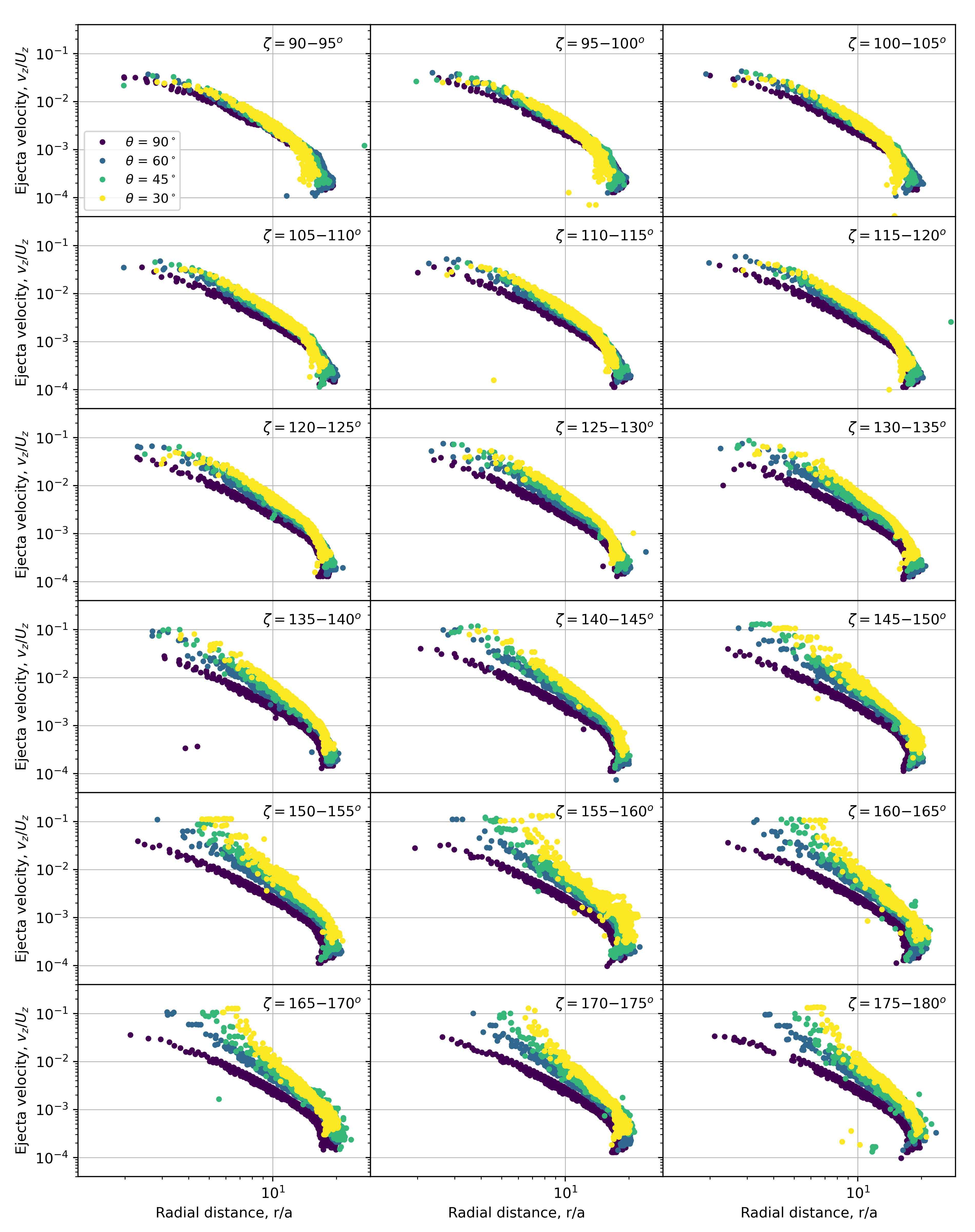}
	\caption{Ejecta vertical velocity - launch position distribution from vertical and oblique impacts ($\theta$ = 90$^\circ$, 60$^\circ$, 45$^\circ$, 30$^\circ$) at azimuth between $\zeta$ = 90 and $\zeta$ = 180$^\circ$.}
	\label{fig:azimuth_fit2}
\end{figure}
\FloatBarrier

Figures~\ref{fig:azimuth_fit} and \ref{fig:azimuth_fit2} show the vertical component of the ejecta velocity normalised by the impact velocity, $v_z/U_z$, as a function of radial distance, normalised by the impactor size $r/a$ and divided into 5$^\circ$ azimuthal segments, for four different impact angles. Here we analyse the vertical component of the ejection velocity as it is most relevant for momentum transfer. In the cross-range direction, at azimuths of about 90$^\circ$, ejection velocity vs launch position is approximately independent of impact angle; however, ejection velocity vs launch position differs systematically with impact angle as azimuth approaches 0$^\circ$ (uprange) or 180$^\circ$ (downrange). In all cases, the middle part of the ejection velocity vs launch position trend is well approximated by a power-law. 

\begin{figure}[!h]
	\centering
	\includegraphics[width=\linewidth]{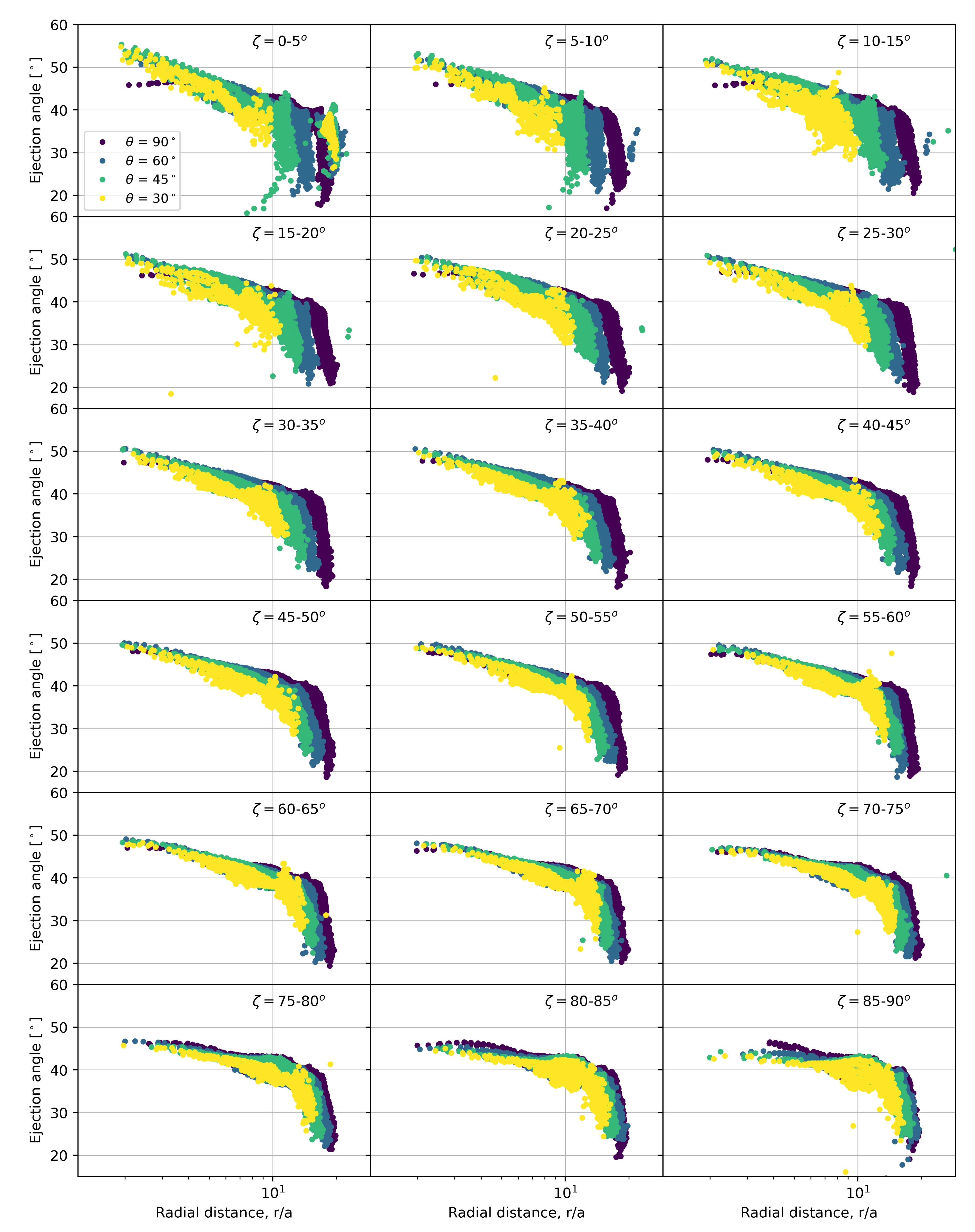}
	\caption{Ejection angle - launch position distribution from vertical and oblique impacts ($\theta$ = 90$^\circ$, 60$^\circ$, 45$^\circ$, 30$^\circ$) at azimuth between $\zeta$ = 0 and $\zeta$ = 90$^\circ$.}
	\label{fig:azimuth_angle}
\end{figure}
\begin{figure}[!h]
	\centering
	\includegraphics[width=\linewidth]{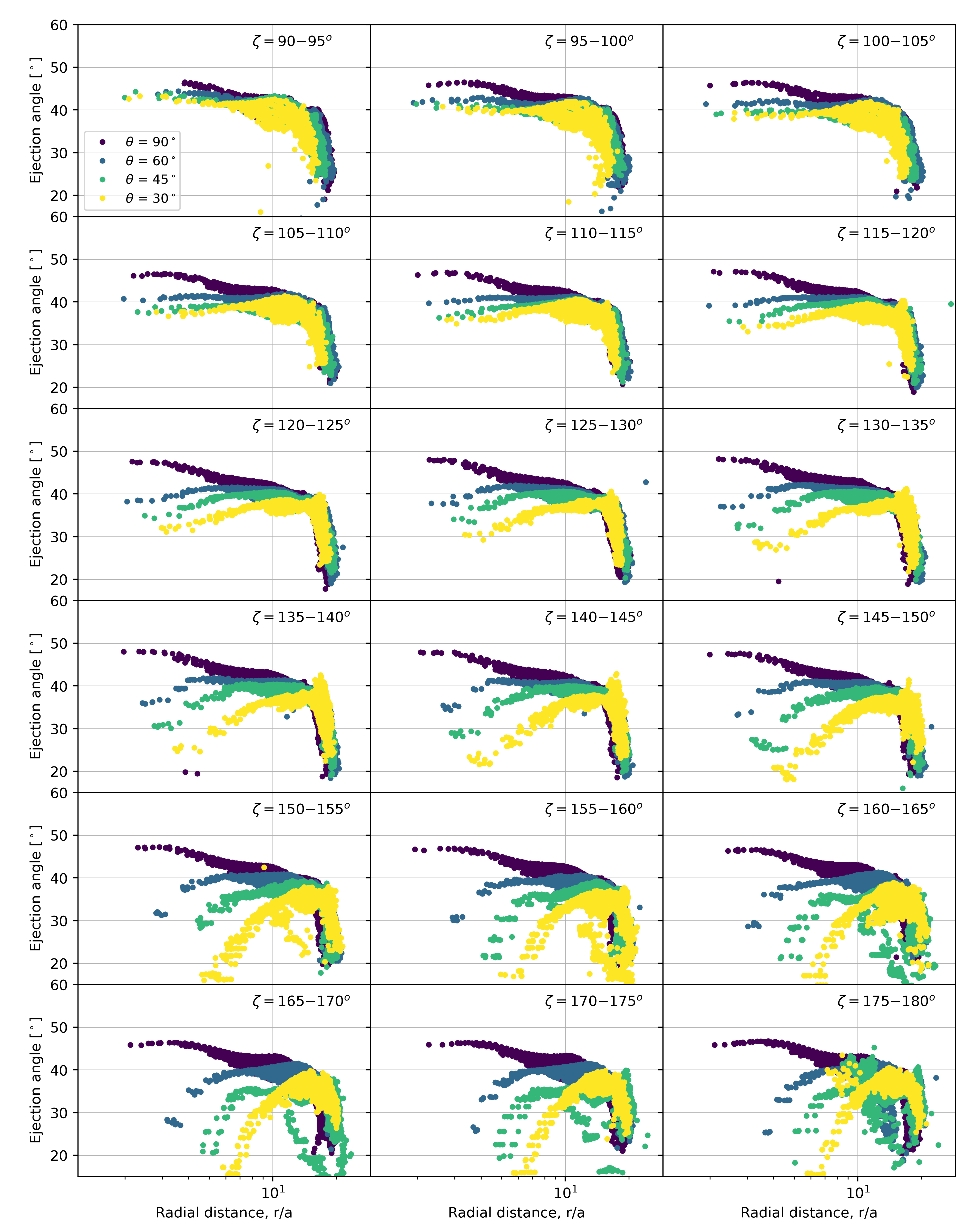}
	\caption{Ejection angle - launch position distribution from vertical and oblique impacts ($\theta$ = 90$^\circ$, 60$^\circ$, 45$^\circ$, 30$^\circ$) at azimuth between $\zeta$ = 90 and $\zeta$ = 180$^\circ$.}
	\label{fig:azimuth_angle2}
\end{figure}

Figures~\ref{fig:azimuth_angle} and \ref{fig:azimuth_angle2} show the ejection angle as a function of radial distance, normalised by the impactor size $r/a$ and divided into 5$^\circ$ azimuthal segments, for four different impact angles. For azimuthal segments between $\zeta = 0^\circ$ and $\zeta = 90^\circ$, the material is ejected at similar ejection angles, regardless of the impact angle (Fig.~\ref{fig:azimuth_angle}). However, between $\zeta = 90^\circ$ and $\zeta = 180^\circ$, the ejection angle decreases with increasing azimuth. At the same time, impacts with lower impact angles, $\theta$, launch material out of the crater at lower ejection angles than in the vertical, $\theta$ = 90 $^\circ$ case. For the four different impact angles, the ejection angle converges close to the crater rim. 

\FloatBarrier

\begin{figure}[!h]
	\centering
	\includegraphics[width=\linewidth]{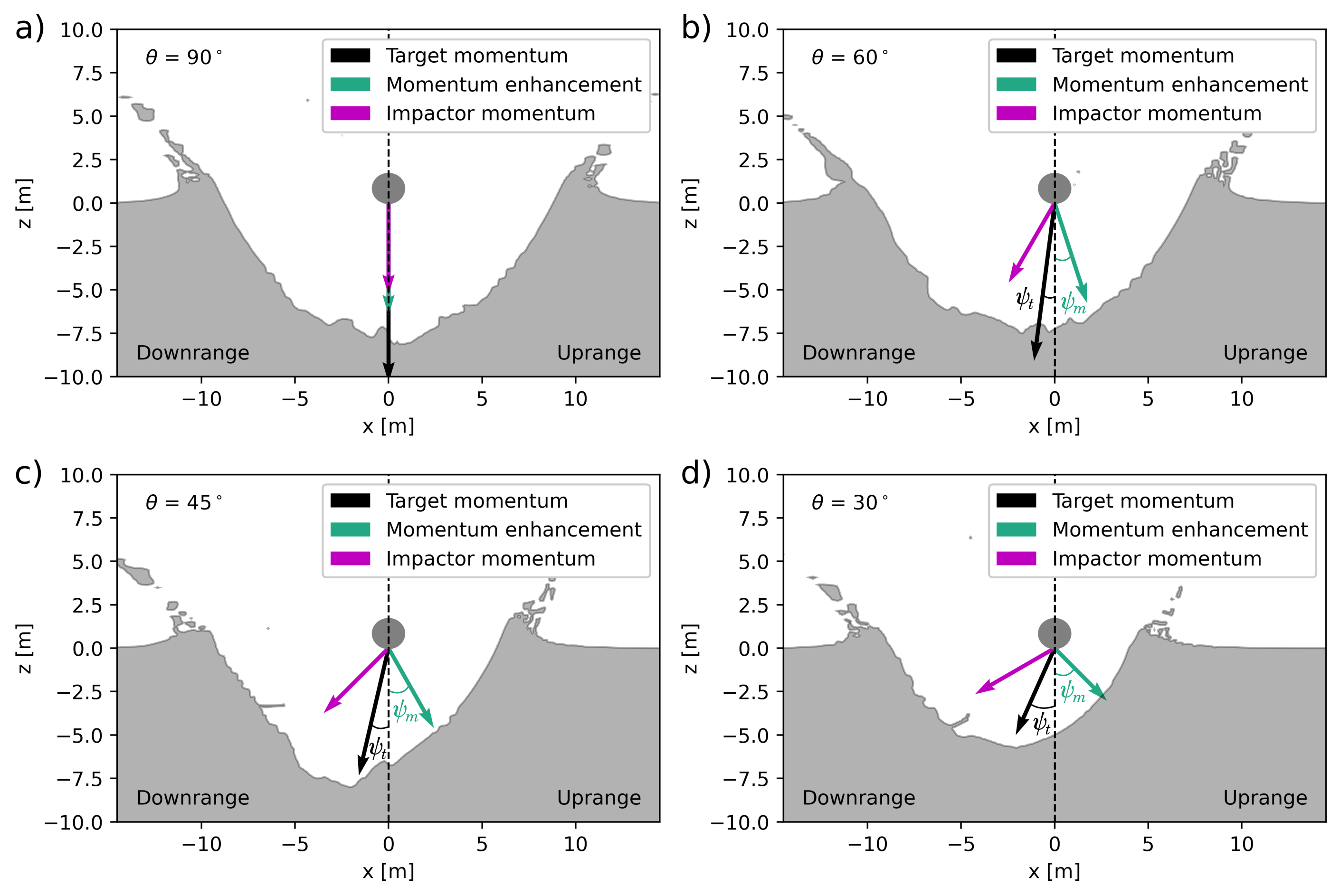}
	\caption{Crater profiles showing the direction of the momentum vectors at the end of crater growth. The impactor momentum is the momentum imparted directly from the impactor, the momentum enhancement is imparted by the ejected particles (makes an angle $\psi_m$ relative to the vertical) and the target momentum is the net momentum of the target, after the impact (makes an angle $\psi_t$ relative to the vertical). The impact direction is right to left.}
	\label{fig:angle_profiles}
\end{figure}

The asymmetry of the ejecta can have important implications for momentum transfer. The net momentum of the target after the impact is the vector sum of the impactor momentum and the net ejected momentum vectors. Figure~\ref{fig:angle_profiles} shows the direction of the momentum vectors for the vertical ($\theta$ = 90$^\circ$) and oblique impacts ($\theta$ = 60$^\circ$, 45$^\circ$ and 30$^\circ$). The projectile imparts an initial momentum along the impact direction, $\mathbf{P_i} = m\mathbf{U}$, where $m$ is the projectile mass and $\mathbf{U}$ is the impact velocity vector. As most of the ejecta momentum is carried away in the downrange direction, the momentum imparted to the target by this ejecta, $\mathbf{P_{ej}}$, is mostly in the uprange direction. We define $\psi_m$ as the angle of the imparted ejecta momentum vector $\mathbf{P_{ej}}$ relative to the vertical. The vector sum of the impactor momentum vector and the imparted ejecta momentum vector is the target momentum, $\mathbf{P_{t}} = \mathbf{P_i} + \mathbf{P_{ej}}$. In the three oblique impact scenarios simulated here, the direction of $\mathbf{P_{t}}$, here defined by the angle $\psi_t$, is between the vertical and the downrange direction. 

\begin{figure}[!h]
	\centering
	\includegraphics[width=\linewidth]{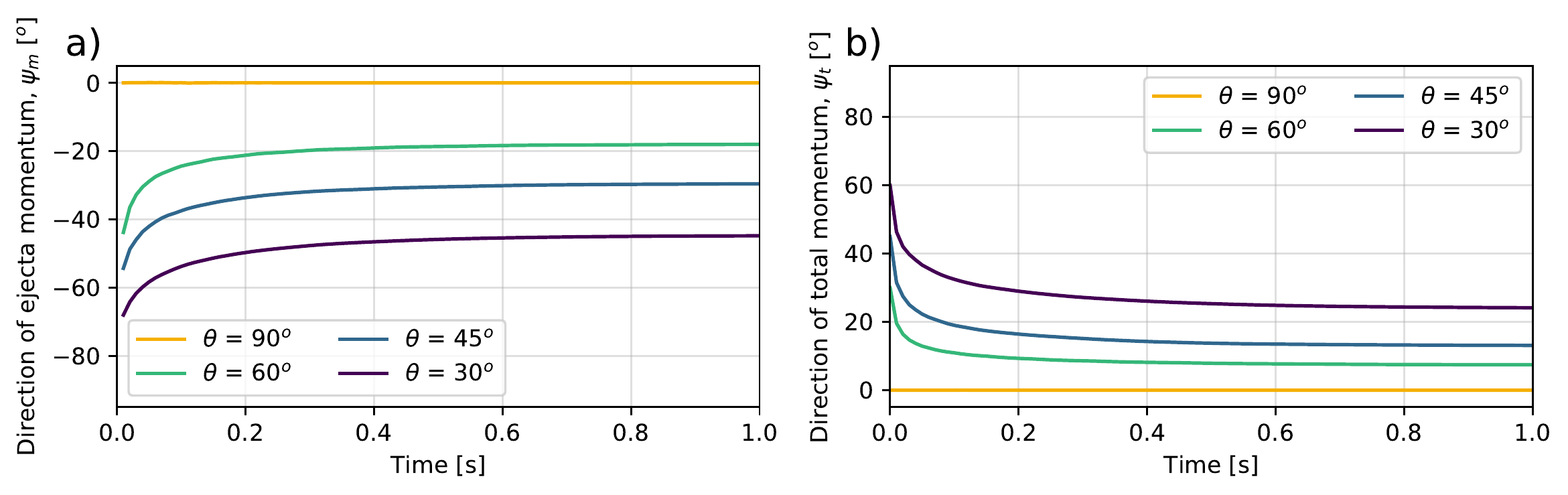}
	\caption{Direction of the (a) ejecta momentum and (b) total momentum from vertical, 90$^\circ$, and oblique, 60$^\circ$, 45$^\circ$ and 30$^\circ$, angle of incidence. The direction is measured anticlockwise from the negative $x$-axis (downrange, 0$^\circ$) to the positive $x$-axis (uprange, 180$^\circ$).}
	\label{fig:beta_time}
\end{figure}

Figure~\ref{fig:beta_time} shows the direction of the momentum imparted by the ejecta, $\psi_m$, (Fig.~\ref{fig:beta_time}a) and the direction of the total momentum, $\psi_t$ (Fig.~\ref{fig:beta_time}b), as a function of time, for the different impact angles.


As the crater grows towards its final diameter, the uprange direction of the ejecta momentum becomes more perpendicular to the surface. The direction of the net momentum imparted on the target also changes, from the downrange direction, towards the vertical direction. In the scenarios simulated here, for impacts into a 10 kPa target, the direction of the net momentum vector at the end of the crater growth is $\psi_m \approx -18^\circ$ for the 60$^\circ$ impact, $\psi_m\approx -30^\circ$ for the 45$^\circ$ impact and $\psi_m\approx -45^\circ$ for the 30$^\circ$ impact.

In the simulations presented here crater growth is halted by the target strength before the ejected momentum direction becomes vertical. However, it is expected that with increasing cratering efficiency (e.g., decreasing strength), the ejecta momentum will make a larger contribution towards the total momentum vector (e.g., the ejecta momentum magnitude will dominate over the impactor momentum magnitude). At the same time, as the cratering efficiency increases, impacts at the same impact angle will become increasingly symmetric and both the ejected and total momentum vectors will be closer to the vertical, $\psi_t \approx \psi_m \approx 0^\circ$. To test this, more numerical simulations of oblique impacts into targets with lower strength are needed, however due to the increased cratering efficiency, such simulations are very computationally expensive. 

\FloatBarrier

\subsection{The effects of target properties on the ejection angle}

\cite{Raducan2019} quantified the effects of target porosity and target coefficient of internal friction on the launch speed of crater ejecta in vertical impacts but the corresponding ejection angles were not reported. Here we present ejection angle results from the same suite of simulations to provide context for our 3D simulation results and to inform analytical approximations of ejecta plume evolution. 

\begin{figure}[!h]
	\centering
	\includegraphics[width=0.5\linewidth]{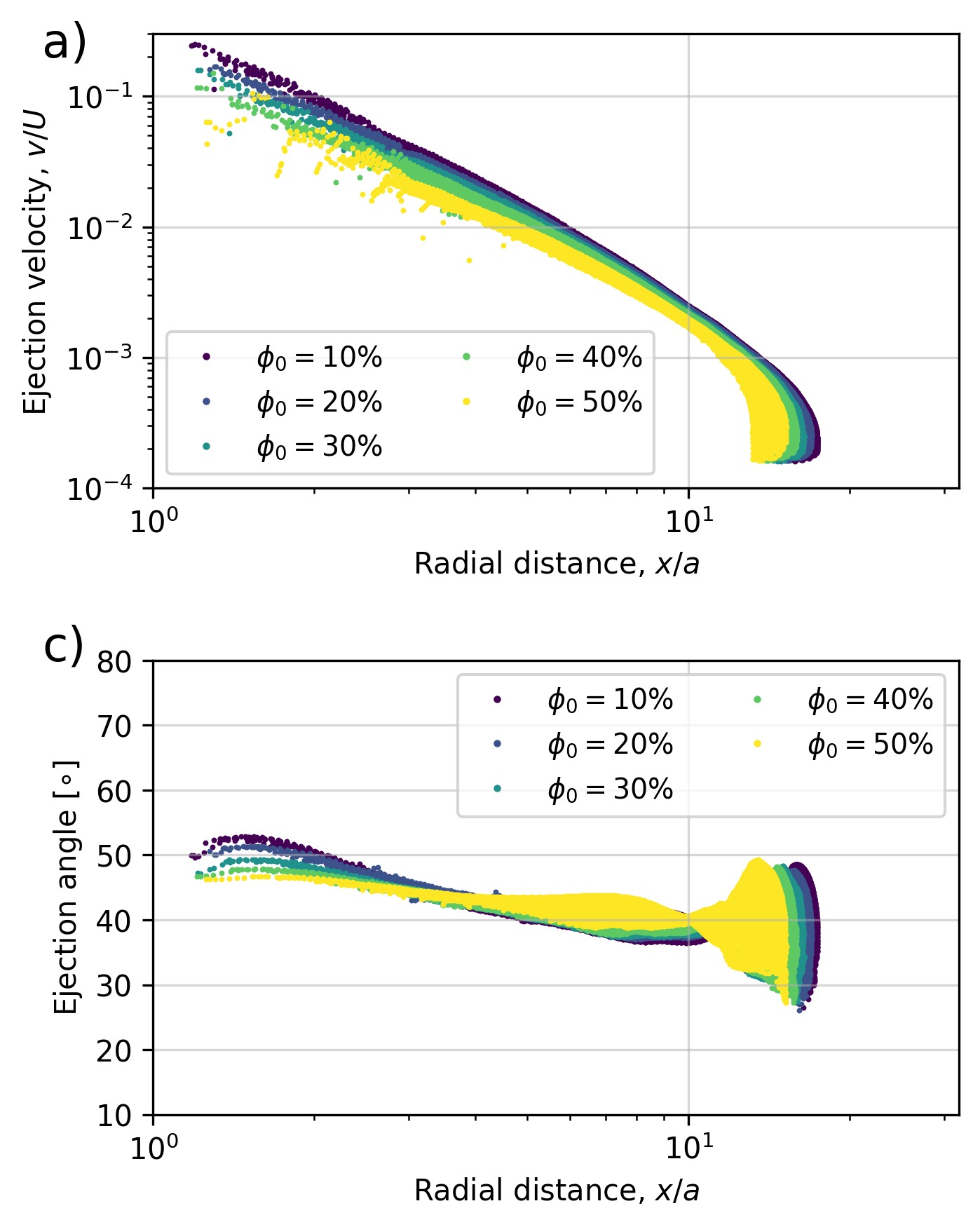}\includegraphics[width=0.5\linewidth]{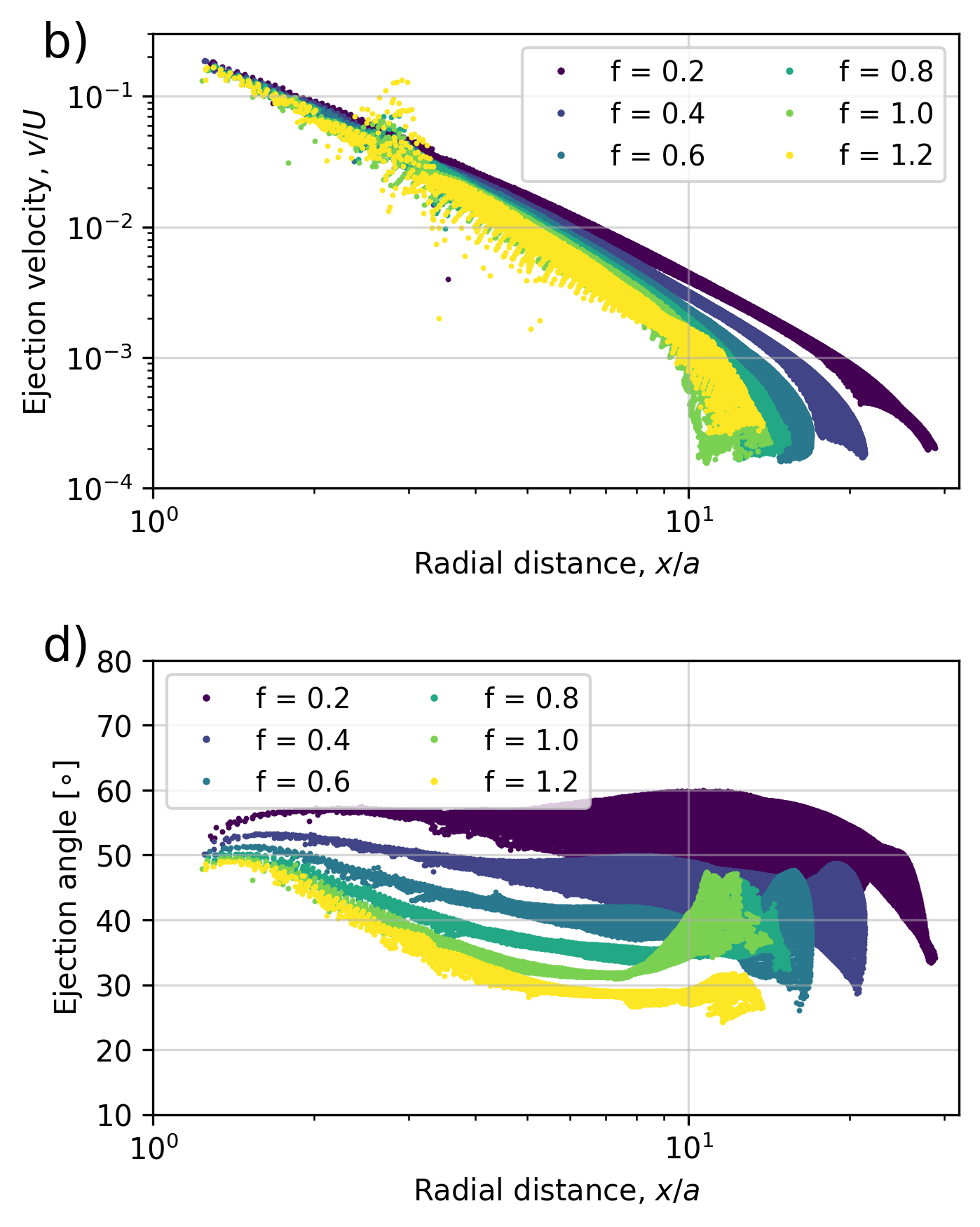}
	\caption{(a, b) Velocity -- launch position distribution of ejecta from impacts into a 10 kPa target, with varying porosity, $\phi_0$ and varying coefficient of friction, $f$. a) represents the ejecta velocity distribution for impacts with $Y_0$ = 10 kPa, $f$ = 0.6 and $\phi_0$ between 10 and 50\%. b) represents the ejecta velocity distribution for impacts with $Y_0$ = 10 kPa, $\phi_0$ = 20\% and $f$ between 0.2 and 1.2. (c, d) Ejection angle -- launch position distribution of ejecta from impacts into a 10 kPa target, with varying $\phi_0$ (c) and varying $f$ (d).} 
	\label{fig:2D_ejecta}
\end{figure}

Figures~\ref{fig:2D_ejecta}(a) and (c) show ejection velocity, normalised by the impact velocity, $v/U$, and ejection angle, respectively, as a function of normalised radial distance from the impact point, $r/a$, for vertical impacts into targets with fixed cohesion, $Y_0$ = 10 kPa, and fixed coefficient of internal friction, $f$ = 0.6. For these targets the initial porosity, $\phi_0$, was varied between 10\% and 50\%. Increasing the target porosity has the effect of reducing the launch speed and ejection angle of the material ejected close to the impact point. 

Figures~\ref{fig:2D_ejecta}(b) and (d) show normalised ejection velocity and ejection angle as a function of normalised radial distance for impacts into 10 kPa, 20\% porous targets, for which the coefficient of internal friction, $f$, was varied between 0.2 and 1.2. Increasing the target coefficient of internal friction has the effect of reducing the launch speed of the material ejected close to the crater rim and the ejection angle. The ejecta velocity-launch position distribution from both parameter studies is discussed in more detail in \citep{Raducan2019}.  

It is often assumed that the ejection angle of individual ejecta particles is approximately 45$^\circ$ to the target surface, independent of launch position. However this assumption does not always hold and the mean ejection angle can vary significantly with time and launch distance, depending on the physical characteristics of the target material, such as porosity or internal friction coefficient \citep{Hoerth2013, Gulde2018, Luther2018}.

Figure~\ref{fig:2D_ejecta}(c) shows that fast ejecta is launched at steeper ejection angles when target porosity is lower. However, the ejection angle converges at radial distances larger than 4$a$ such that the ejection angle of slower ejecta is relatively insensitive to target porosity. 

Figure~\ref{fig:2D_ejecta}(d) shows that ejection angle decreases with launch position, by up to 20$^\circ$. At the same time, the average ejection angle decreases with increasing coefficient of internal friction, from about 60$^\circ$ for $f$ = 0.2, to about 30$^\circ$ for $f$ = 1.2. Similar trends have been observed for vertical impacts in the gravity regime \citep{Luther2018}. 

\section{Discussion}

\subsection{Towards an ejecta scaling relationship for oblique impacts}

Ejecta scaling relationships are useful to determine the ejecta mass--velocity distribution and momentum transfer for vertical impacts scenarios other than the ones considered here. However, most planetary impacts are oblique and the existing scaling relationships \citep{Housen2011} only apply to vertical impacts. Therefore, the current point-source scaling theory needs to be extended and adapted to oblique impacts. 

Previous attempts at determining the ejecta distribution from oblique impacts empirically include the studies by \citet{Anderson2003}, \citet{Anderson2004} and \citet{Richardson2007}. \citet{Anderson2003} conducted impact experiments of $\approx$ 6 mm diameter aluminium spheres into medium-grained sand, at 90$^\circ$ and 30$^\circ$ impact angles and an impact velocity of 1 km/s. From each impact experiment they recorded the ejection velocities, angles, and positions of the ejecta expelled at one moment during the first half of the crater growth. They observed that in the oblique impact cases, the velocity distribution of the ejected particles varies with azimuth from the impact point relative to the trajectory direction. They defined the ratio between the downrange and the uprange ejection velocities (DR/UR) as a measure of the asymmetry in the ejecta curtain. At early times, they recorded a 40\% increase in the ejection velocity from the uprange to the downrange sides of the crater. At later times, about when the crater radius reached about half of the final radius, the difference between the velocities in the uprange and in the downrange sides of the craters decreased to about 20\%. \citet{Anderson2004} ran additional impact experiments at 45$^\circ$ impact angles and tried to use the Maxwell Z-Model to predict the ejection velocities and ejection angles. One major difficulty with deriving a scaling law using this approach was the assumption that there is a single, stationary point source. 

The work in this section attempts to develop an ejecta scaling relationship for oblique impacts, based on numerical simulation data. The three-dimensional simulations of the DART impact into a 10 kPa, 20\% porous target, at vertical, 60$^\circ$, 45$^\circ$ and 30$^\circ$ impact angles presented above are used here to provide information about the ejecta distribution as a function of impact angle. 

Fitting Eq.~\eqref{eq:v_x} through the ejection velocity-launch position distributions shown in Figures~\ref{fig:azimuth_fit} and \ref{fig:azimuth_fit2} (where U is the vertical component of the impact velocity, $U_z$), allows us to determine $\mu$ and $C_1$ for each azimuth and impact angle investigated. The azimuthal radius, $R_{\zeta}$, which here is defined as the distance between the impact point and the crater rim for each azimuthal segment can also be determined. Similarly, the constant $k$ can be found by fitting Eq.\eqref{eq:m_x} to the cumulative ejected mass-launch position data. 

\begin{figure}[!h]
	\centering
	\includegraphics[width=\linewidth]{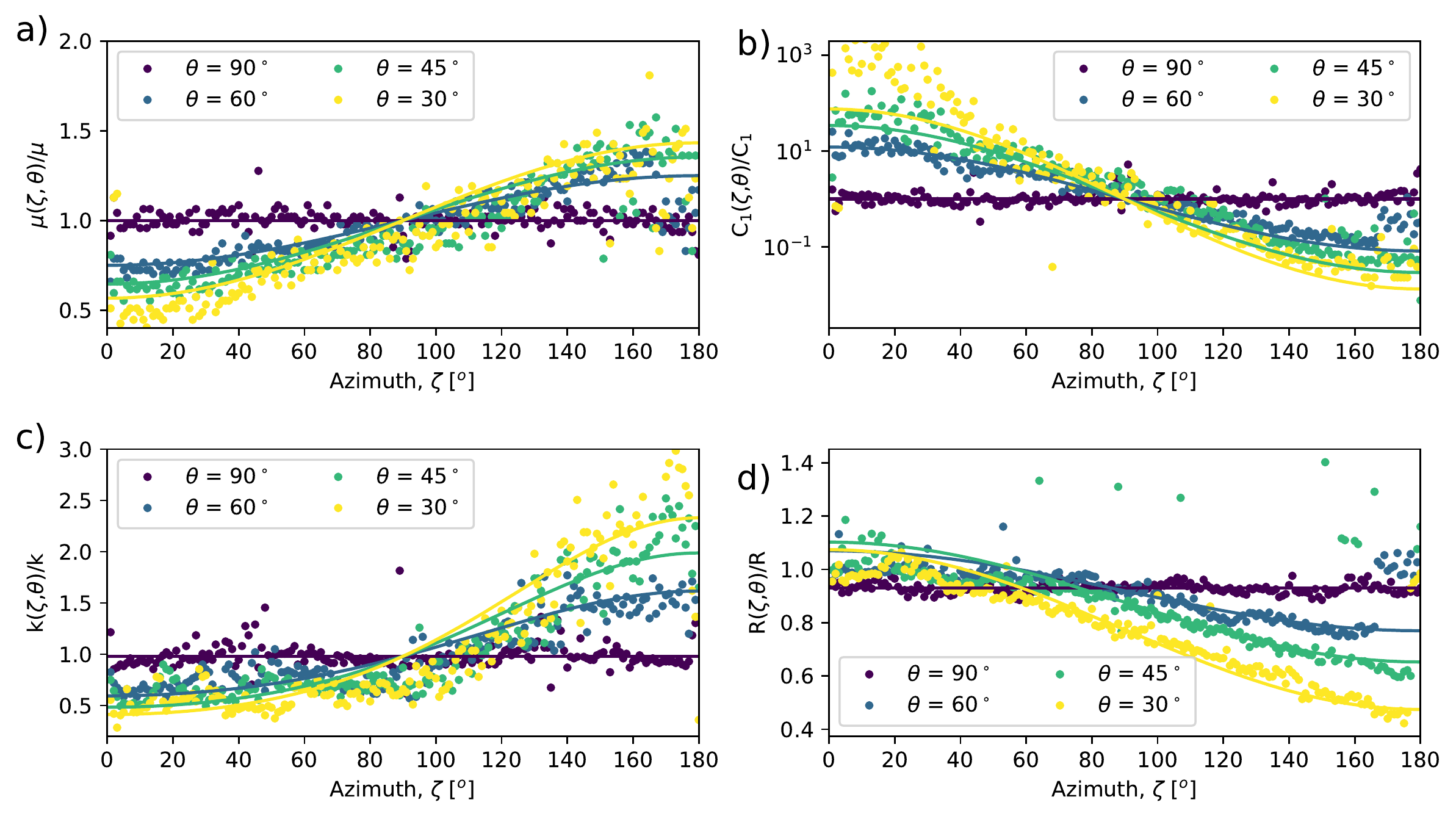}
	\caption{Constants (a) $\mu$, (b) $C_1$, (c) $k$ and (d) $R$, normalised by the average value in the 90$^\circ$ impact scenario, as a function of azimuth around the crater (centred at the impact point).}
	\label{fig:law_y}
\end{figure}

Figure~\ref{fig:law_y} shows the best fit constants $\mu$ (Fig.~\ref{fig:law_y}a), $C_1$ (Fig.~\ref{fig:law_y}b), $k$ (Fig.~\ref{fig:law_y}c) and $R_{\zeta}$ (Fig.~\ref{fig:law_y}d) as a function of azimuth, $\zeta$, for the four different impact angles. The constants were normalised by the average $\mu$, $C_1$, $k$ and $R$ constants found for the $\theta$ = 90$^\circ$ impact scenario. All four constants, $\mu$, $C_1$, $k$ and $R_{\zeta}$, vary as periodic functions of impact angle and azimuth around the impact point. In this work, simple trigonometric functions are fit through these distributions as a function of azimuth for each impact angle.  

The coupling parameter velocity exponent, $\mu(\zeta, \theta)$, can be approximated by 
\begin{equation}\label{eq:mu_az}
    \mu(\zeta, \theta) \approx \mu\times\left[1 +\frac{1}{2}\cos(\zeta)\cos(\theta) \right],
\end{equation}
where $\mu$ is the velocity exponent in the vertical impact scenario. 
The constant $C_1(\zeta, \theta)$ varies as an exponential of cosine
\begin{equation}\label{eq:c1_az}
    C_1(\zeta, \theta) \approx C_1 \times \exp{\left[-5\cos(\zeta)\cos(\theta)\right]},
\end{equation}
where $C_1$ is the constant derived for the vertical impact scenario. 
\begin{equation}\label{eq:k_az}
    k(\zeta, \theta) \approx \frac{k}{n_{\zeta}} \times \exp{\left[-0.02\cos(\zeta)\cos(\theta)\right]},
\end{equation}
where $k$ is the constant derived for the vertical impact scenario and $n_{\zeta}$ is the number of azimuthal segments considered.  
\begin{equation}\label{eq:R_az}
    R(\zeta, \theta) \approx R \times \left[1-\frac{(90-\theta)}{100}\times\frac{\cos(\zeta)}{2} \right],
\end{equation}
where $R$ is the crater radius in the vertical impact scenario. These trigonometric functions were plotted in Figure~\ref{fig:law_y}.

\subsection{Comparison of the oblique impact ejecta scaling relationship with laboratory experiments}

Equations~\eqref{eq:mu_az}--\eqref{eq:R_az}, can be used together with Eqs.~\eqref{eq:v_x}--\eqref{eq:m_x} to determine the ejection velocity as a function of launch position and azimuth for an oblique impact, given known scaling constants ($\mu$, $C_1$, $k$, $R$). Here we present a methodology for using these equations to approximate the distribution of ejection velocities from an oblique impact, at fixed times during crater growth, using only information that can be derived from an equivalent vertical impact. To test our methodology, we used the experimental ejection velocity data recorded at several different radial distances from the impact point for laboratory-scale impacts into sand at a speed of $\approx$ 1 km/s and at impact angles of both 30$^\circ$ and 90$^\circ$ to the horizontal \citep{Anderson2003}. 

\begin{figure}[!h]
	\centering
	\includegraphics[width=0.8\linewidth]{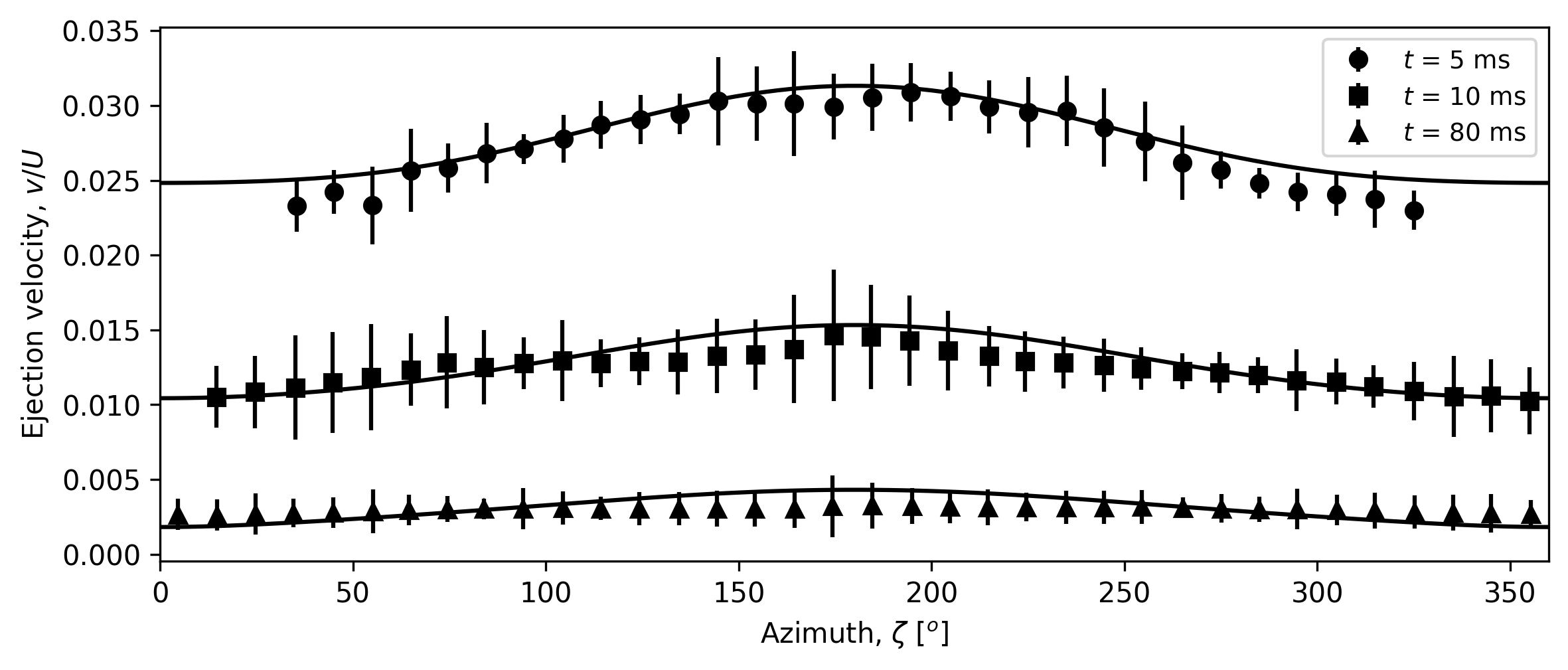}
	\caption{Average ejection velocity from $\approx$ 1 km/s oblique 30$^\circ$ impacts into sand conducted in laboratory by \citep{Anderson2003}, as a function of azimuth around the crater, at thee different times: $t$ = 5, 10 and 80 ms. The error bars were calculated as 1$\sigma$ \citep{Richardson2007}. Our semi-analytical approximation of the ejection velocity from Eqs.~\eqref{eq:mu_az}--\eqref{eq:c1_az} and Eq.~\eqref{eq:v_x2} was plotted for comparison.}
	\label{fig:ejecta_exp}
\end{figure}

\begin{figure}[!h]
	\centering
	\includegraphics[width=0.6\linewidth]{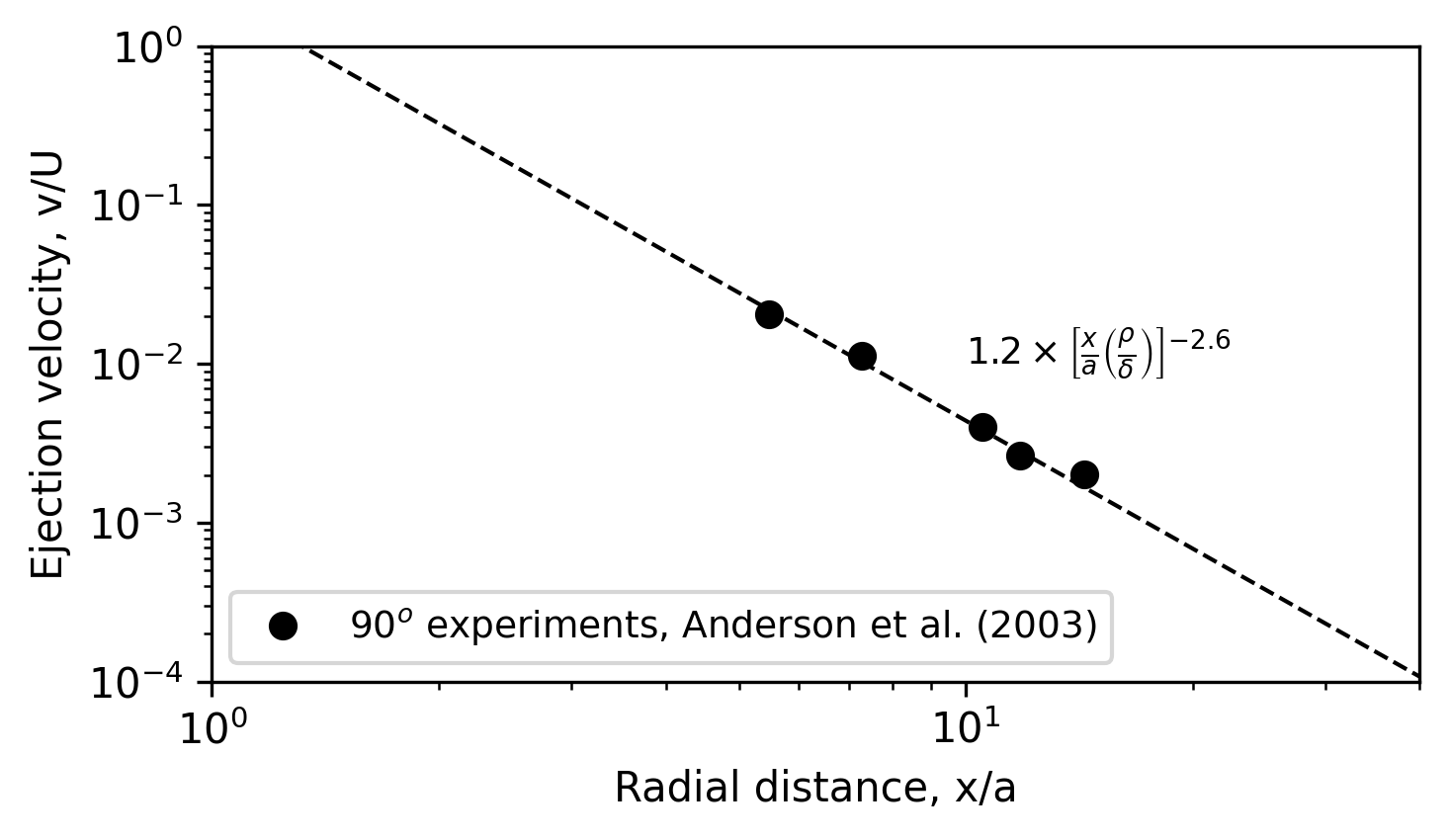}
	\caption{Ejection velocity vs radial distance distribution of ejecta from $\approx$ 1 km/s vertical impacts \citep{Anderson2003}.  }
	\label{fig:and_vel90}
\end{figure}

The first step is to derive the scaling constants $\mu$ and $C_1$ from ejection velocity data for a vertical impact into the same target material and otherwise similar impactor parameters to the oblique impacts in question. The vertical impact experiments of \cite{Anderson2003} are presented in the form of $v/(gR)^{1/2}$ versus $x/R$, where $v$ is ejection speed, $R$ is the apparent crater radius and $x$ is the launch position. Ejection velocity and  radial distance were therefore converted into $v/U$ and $x/a$ values using an apparent crater radius of $R$ = 8.1\,cm (see \cite{Housen2011} for details), where $a$ = 3.175\,mm (Fig.~\ref{fig:and_vel90}). By fitting Eq.~\eqref{eq:v_x} to the rescaled experimental data, we derived $\mu$ = 0.38, $C_1$ = 1.23.

The next step is to substitute these derived constants into Eqs.~\eqref{eq:mu_az}--\eqref{eq:c1_az}, which can then be substituted into Eq.~\eqref{eq:v_x2}. Here we used $\nu$ = 0.4, $n_1$ = 1, $n_2$ = 1.2, $p$ = 0.3, as derived by \cite{Housen2011} for impacts into sand, and $q$ = 0.2 \citep{Raducan2019}. In these experiments, the crater radius was measured as $R$ = 8.1\,cm \citep{Housen2011}, however the crater radius can also be approximated from Eq.~\eqref{eq:pi_2} (or Eq.~\eqref{eq:pi_3} for the strength regime). 

Figure~\ref{fig:ejecta_exp} shows the average ejection velocity as a function of azimuth, as measured by \cite{Anderson2003}, at 5, 10 and 80 ms after impact.  For comparison, our analytical approximation of the ejection velocity (Eq.~\eqref{eq:v_x2}) from a 30$^\circ$ impact as a function of azimuth, $\zeta$, is plotted as a continuous line. To convert from radial launch position to ejection time, we follow the approach of \cite{Richardson2007}: at 5 ms, $x/a \approx 0.2$; at 30 ms, $x/a \approx 0.3$; at 80 ms, $x/a \approx 0.5$.

Our semi-analytical model of ejection velocities shows a generally good agreement with the experimental data. The match is least impressive at late times, close to the crater rim. The laboratory experiments presented here have a much higher cratering efficiency than the the numerical simulations used to derive the trigonometric functions used here, which might be the source of the discrepancy. Therefore, for cratering events with much larger cratering efficiencies than our simulations one should apply caution when using our approach to approximate the velocity distribution of the slow ejecta from oblique impacts. 

\FloatBarrier
\subsection{Semi-analytical approximation for the momentum enhancement from oblique impacts}

The efficiency of impact momentum transfer is often expressed in terms of a factor $\beta$, which for a vertical impact is simply the ratio of the momentum transferred to the target $M\Delta v$ (where $M$ is the target mass and $\Delta v$ is the velocity change) divided by the impactor momentum $mU$. More generally, the conservation of momentum can be defined as:

\begin{equation}
    M\mathbf{\Delta v} = m\mathbf{U} + m(\beta-1)(\mathbf{\hat{n}}\cdot\mathbf{U})\mathbf{\hat{n}} + m(\gamma-1)(\mathbf{\hat{t}}\cdot\mathbf{U})\mathbf{\hat{t}} 
    \label{eq:momentum}
\end{equation}

\noindent where $\mathbf{\hat{n}}$ and $\mathbf{\hat{t}}$ are the inward surface normal, and downrange directed surface tangent unit vectors, respectively. In our simulations of impacts into a flat target, with the impact trajectory in the $x$-$z$ plane, the inward surface normal vector is in the negative vertical direction $-z$ and the downrange surface tangent vector is $-x$. The first term on the right-hand side is the momentum imparted directly by the impactor $\mathbf{P_i}$; the second term is the component of the momentum imparted by the escaping ejecta that acts normal to the surface; the third term is the component of the momentum imparted by the escaping ejecta that acts parallel to the surface, positive in the downrange direction. If the net momentum imparted to the target by the escaping ejecta can be assumed to act along the surface normal vector (i.e., $\psi_m = 0$) then the third term can be neglected \citep{Feldhacker2017, Cheng2020}. However, for the scenarios simulated here $\lvert\psi_m\rvert > 0$ and so we retain the full expression.

According to Equation~\ref{eq:momentum}, the definition of $\beta$ (and $\gamma$) can be expressed in terms of the normal (and tangential) components of the momentum transfer and incident moment:

\begin{eqnarray}
\beta & = & \frac{M \mathbf{\hat{n}}\cdot\mathbf{\Delta v} }{m \mathbf{\hat{n}}\cdot\mathbf{U}} \ = \ \frac{M\Delta v \cos\psi_t}{mU \sin\theta},\\
\gamma & = & \frac{M \mathbf{\hat{t}}\cdot\mathbf{\Delta v} }{m \mathbf{\hat{t}}\cdot\mathbf{U}} \ = \ \frac{M\Delta v \sin\psi_t}{mU \cos\theta} .
\end{eqnarray}

\noindent Alternatively, $\beta$ and $\gamma$ can be expressed in terms of the normal and tangential components of the net ejecta momentum $\mathbf{P_{ej}}$:

\begin{eqnarray}
\beta - 1 & = & \frac{\mathbf{\hat{n}}\cdot\mathbf{P_{ej}} }{m \mathbf{\hat{n}}\cdot\mathbf{U}} \ = \ \frac{P_{ej} \cos\psi_m}{mU \sin\theta},\\
\gamma - 1 & = & \frac{\mathbf{\hat{t}}\cdot\mathbf{P_{ej}} }{m \mathbf{\hat{t}}\cdot\mathbf{U}} \ = \ \frac{P_{ej} \sin\psi_m}{mU \cos\theta} .
\end{eqnarray}

\noindent Hence, $\beta$ and $\gamma$ are related by:

\begin{equation}
    (\gamma-1) = (\beta-1)\tan \theta \tan \psi_m
    \label{eq:gamma}
\end{equation}

\noindent for $\theta < 90$ and $\lvert\psi_m\rvert < 90$. Note that for vertical impacts $\gamma = 0$ and $(\gamma-1) = 0$ if the net ejecta momentum is directed exactly normal to the target ($\psi_m = 0$).

For a vertical impact ($\theta = 90$, $\psi_m = 0$), \citet{Cheng2016} showed that the vertical momentum carried away by the ejecta from a vertical impact, $\beta-1$, can be found from integrating the differential mass, $dM$ (from Eq.~\eqref{eq:m_x}), within the radial distance range from $n_1$ to $n_2R/a$ 
\begin{equation}
   P_{ej} =  \frac{9k m}{4\pi} \frac{\rho}{\delta}  \int_{n_1}^{\frac{n_2R}{a}} w^2 v_z dw,
\end{equation}
where $w$ = $r/a$ and $v_z$ is the vertical component of the ejection velocity. The vertical momentum carried away by the ejecta, $\beta-1$ is then:

\begin{equation}
    \beta-1 = \frac{P_{ej}}{mU} = \frac{9k}{4\pi} \frac{\rho}{\delta}  \int_{n_1}^{\frac{n_2R}{a}} w^2 \frac{v_z}{U} dw.
\end{equation}

To generalise this approach for oblique impacts, we define the vertical ejecta momentum, $\beta-1$, as the sum of the momentum calculated for each azimuth wedge:
\begin{equation}\label{eq:beta_sum1}
    \beta-1 \ = \ \frac{\mathbf{\hat{n}}\cdot\mathbf{P_{ej}} }{m \mathbf{\hat{n}}\cdot\mathbf{U}} \ = \ \frac{P_{ej} \cos\psi_m}{mU \sin\theta}\approx 2\times \sum_{i=1}^{n_\zeta} \frac{9k_{\zeta}}{4\pi} \frac{\rho}{\delta}  \int_{R_{\zeta}}^{n_1a} \frac{v_z(w)}{U_z}w^2 dw
\end{equation}
where $w$ = $r/a$, $k_{\zeta} = k/n_{\zeta}$ and $n_\zeta$ is the number of azimuthal wedges between $\zeta=0$ (uprange) and $\zeta=180$ (downrange). To evaluate the integral requires that we substitute the power-law approximation of Eq.~\eqref{eq:v_x} with parameters appropriate for each azimuthal wedge into Eq.~\eqref{eq:beta_sum1}, noting that the relevant normalised ejection velocity component is $v_z/U_z$. In this case, the momentum enhancement can be approximated by
\begin{equation}\label{eq:beta_sum}
    \beta-1 \approx 2\times\sum_{i=1}^{n_\zeta} \frac{9k(\zeta_i,\theta)}{4\pi \sin \theta} \frac{\rho}{\delta} \int_{n_1}^{\frac{n_2R(\zeta_i, \theta)}{a}}  \left[C_1(\zeta_i,\theta)\left[\left(w\right)\left(\frac{\rho}{\delta}\right)^\nu\right]^{-1/\mu(\zeta_i,\theta)} \right]w^2 dw.
\end{equation}
To calculate $\beta-1$ for a given set of constants, Eq.~\eqref{eq:beta_sum} must be solved numerically. 

Equation~\eqref{eq:beta_sum} provides a framework for estimating $\beta-1$ for an impact with known impactor ($m$, $U$, $\delta$, $\theta$) and target parameters ($\rho$). It requires knowledge of the azimuthal and angle dependence of the ejecta distribution parameters $k(\zeta,\theta)$, $C_1(\zeta,\theta)$ and $\mu(\zeta,\theta)$ as well as the azimuthal and angle dependence of the crater rim radius, relative to the impact point $R(\zeta,\theta)$. With knowledge of the ejecta imparted momentum vector angle to the vertical, $\psi_m$, $\gamma-1$ can be determined from $\beta-1$ using Eq.~\ref{eq:gamma}.

Equations~\eqref{eq:mu_az}-\eqref{eq:R_az} approximate these relationships for impacts similar to those simulated here and may be more broadly applicable given the promising comparison with experimental data. However, the required values of these parameters can also be determined for each azimuth and impact angle from individual simulation results. To measure the error in each level of approximation, we compare the total integrated vertical ejected momentum for each impact angle, as determined by summing the contribution of each ejected tracer particle, with two estimates of the total ejected momentum. In Table~\ref{table:best_fit_beta} the total vertical momentum of the ejecta as derived from 
summing the contribution of each ejected tracer particle from each simulation is defined as `Measured $\beta-1$'.  
The first estimate of the ejecta vertical momentum (`Best-fit $\beta-1$' in Table~\ref{table:best_fit_beta}) comes from using Eq.~\eqref{eq:beta_sum} and individual analytical best-fit constants ($\mu$, $C_1$, $k$ and $R_{\zeta}$) for each azimuth as shown in Fig.~\ref{fig:law_y}. The second estimate  (`Analytical $\beta-1$' in Table~\ref{table:best_fit_beta}) was obtained from using Eq.~\eqref{eq:beta_sum} and the analytical expressions for: $\mu(\zeta,\theta)$ (Eq.~\eqref{eq:mu_az}), $C_1(\zeta,\theta)$ (Eq.~\eqref{eq:c1_az}),  $k(\zeta,\theta)$ (Eq.~\eqref{eq:k_az}) and $R(\zeta,\theta)$ (Eq.~\eqref{eq:R_az}). \cite{Raducan2019} determined that for a vertical impact into a 1 kPa, 20\% porous target the fitting constants are $\mu \approx$ 0.42, $C_1 \approx$ 1.1 and $k \approx$ 0.4. Here we used Eq.~\eqref{eq:beta_sum} to find an analytical approximation of $\beta-1$ for this set of scaling constants. Our comparison between the three different measures of $\beta-1$ show that both estimates are within 2\% of the `measured' $\beta-1$ for a 90$^\circ$ impact and within 20\% for a 30$^\circ$ impact.

\begin{table}[ht]
\caption{Normal ejecta momentum, $\beta-1$, and tangential ejecta momentum, $\gamma-1$, measured from the Lagrangian tracers, compared with $\beta-1$ calculated from the best-fit constants and calculated using the newly derived analytical approximations (Eq.~\eqref{eq:beta_sum}). }
\label{table:best_fit_beta} 
\begin{center}
	\begin{tabular}{l@{\hskip 0.5in}c@{\hskip 0.5in}c@{\hskip 0.5in}c@{\hskip 0.5in}c}
		\hline
		& \multicolumn{4}{c}{Impact angle}\\
		& 90$^{\circ}$ & 60$^{\circ}$ & 45$^{\circ}$ & 30$^{\circ}$ \\
		\hline
        Measured $\beta-1$ & 1.32 & 1.36 & 1.33 & 1.24 \\
        Direction ejecta momentum, $\psi_m$ & 0.03$^\circ$ & -18.00$^\circ$ & -29.57$^\circ$ & -44.79$^\circ$ \\
        Direction total momentum, $\psi_t$ & 0.02$^\circ$ &  7.42$^\circ$ & 13.09$^\circ$ & 24.13$^\circ$ \\
        Measured $\gamma-1$ & 0.00 & -0.38 & -0.53 & -0.61 \\
        Best-fit $\beta-1$ & 1.24 & 1.35 & 1.37 & 1.02 \\
        Analytical $\beta-1$ & 1.30 & 1.35 & 1.36 & 1.52\\
		\hline
	\end{tabular}
\end{center}
\end{table}

For the impact scenarios studied here, the vertical efficiency of the momentum transfer seems to be almost constant, with less than 4\% variation between $\beta-1$ at 90$^\circ$ and $\beta-1$ at 30$^\circ$ impact angle. 

\subsection{Implications for the DART and Hera missions}

The change in momentum caused by the DART impact will be approximated from the circular Keplerian motion of the Didymos system \citep{Cheng2020} and therefore only the momentum transfer in the direction of the target's orbital velocity will be measured by the DART mission. In an ideal scenario, the DART impactor would strike Dimorphos's surface vertically (normal incidence; $\theta=90$) in the direction of its orbital velocity. In an oblique impact, due to the asymmetric distribution of the crater ejecta, the change in momentum is not co-linear with the impactor momentum. If the DART impact occurs at a non-vertical impact angle $\theta < 90$ then the momentum change in the orbital velocity direction can be defined as $M\Delta v/mU = \beta\sin\theta/\cos\psi$, which for the impact scenarios considered here can be reasonably approximated as $M\Delta v/mU = \beta\sin\theta$.

As as seen from previous studies, the efficiency of the vertical momentum transfer produced by the DART impact, $\beta\sin\theta$, is very sensitive to the target properties and impact conditions \citep{Jutzi2014, Stickle2015, Syal2016, Raducan2019, Raducan2020}. Moreover, a number of target property configurations (e.g., different target cohesion-porosity-impact angle combinations) can result in the same deflection \citep{Raducan2020}. Therefore an observed value of $\beta$ can be interpreted in different ways depending on the target and impact properties, which will not be known before the arrival of the Hera mission. 

Having demonstrated that the analytical expressions derived here give reasonable approximations of $\beta-1$, which are within 20\% of the numerical data, the same methodology can then be used to extrapolate the $\beta$ values for impacts into targets with cohesions and porosities different to those used in the oblique impact simulations presented here. 

Of principal interest for the DART mission is the normal momentum transfer, $M\Delta v/mU \approx \beta\sin\theta\cos\psi_t$.
\cite{Raducan2019} gives values for $\mu$, $C_1$ and $k$ for DART-like vertical impacts into asteroid targets with various cohesions, porosities and internal friction coefficients. Substituting the $\mu$, $C_1$ and $k$ constants derived from two-dimensional vertical impact simulations \citep{Raducan2019} into Eq.~\eqref{eq:beta_sum}, $M\Delta v/mU$ can be calculated for a range of target cohesions and fixed porosity. The crater radius, $R$, was calculated using Eq.~\eqref{eq:pi_3}.

\begin{figure}[!h]
	\centering 
	\includegraphics[width=\linewidth]{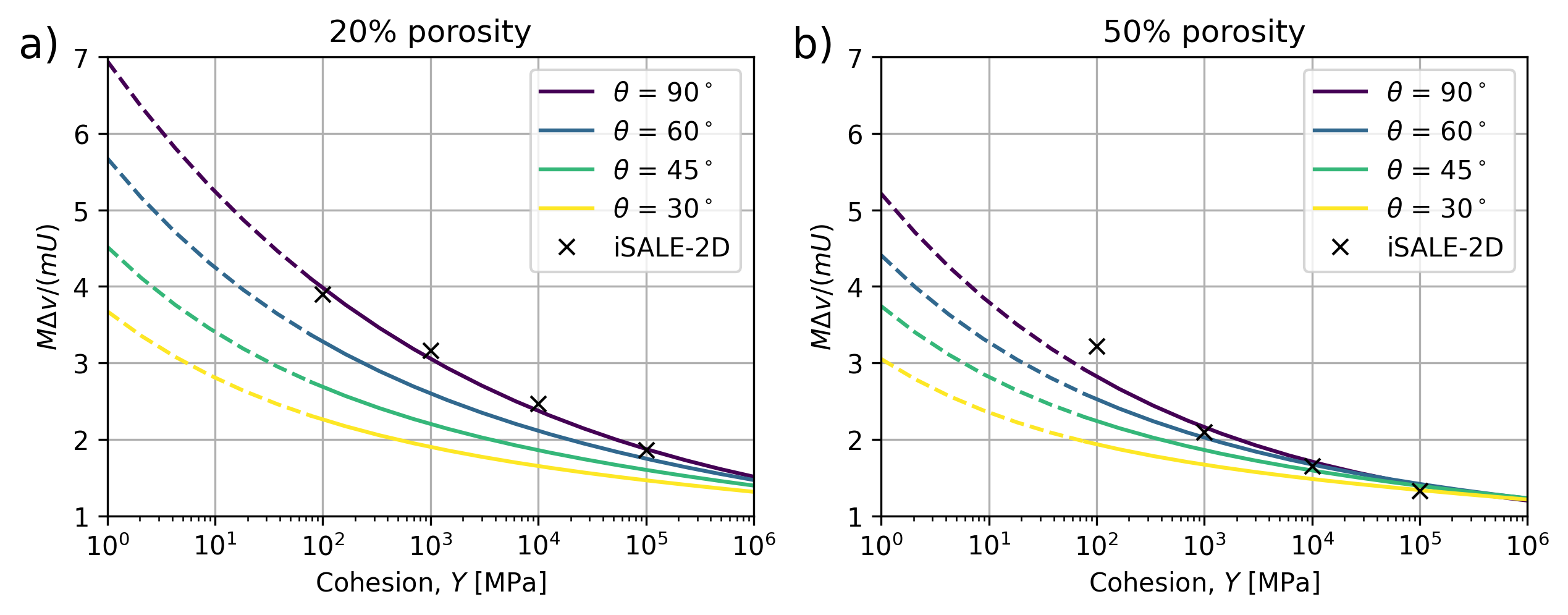}
	\caption{$M\Delta v/mU$ trends found using Eq.~\eqref{eq:beta_sum}, for oblique impacts into a 20\% porous and a 50\% porous target.}
	\label{fig:law_20_50}
\end{figure}
\FloatBarrier

Figure~\ref{fig:law_20_50} shows the analytical $M\Delta v/mU (\approx \beta\sin\theta)$ estimates for impacts at vertical, 60$^\circ$, 45$^\circ$ and 30$^\circ$, into targets with a cohesion between 1 Pa and 1 MPa and a fixed porosity of 20\% (Fig.~\ref{fig:law_20_50}a) and 50\% (Fig.~\ref{fig:law_20_50}b). This analytical approximation shows that for a 100\,Pa, 20\% porous target, a 30$^\circ$ impact angle would reduce $M\Delta v/mU$ by up to 50\%. This relative difference seems to increase with decreasing cohesion. On the other hand, for a 100\,Pa, 50\% porous target, a 30$^\circ$ impact angle would reduce $M\Delta v/mU$ by about 30\%. Further impact simulations at oblique angles into targets with a range of cohesions and porosities will verify and extend the limits of applicability of these scaling relationships and quantify any associated errors.  

The semi-analytical model presented in this work and shown in Figure~\ref{fig:law_20_50} provides a framework to determine the expected $\beta-1$ values from an oblique impact, given known target properties and impact conditions. However this analysis can also work in reverse: from known $\beta-1$ values, one could determine the range of possible target properties and impact conditions. In this way, the $\beta$ value calculated from the observed change in the asteroid's orbit can be used to determine the range of possible target properties that would produce such deflection. An advantage of this semi-empirical approach is that it avoids the need for computationally expensive numerical simulations to span a very large parameter space, especially in three-dimensions.  

In addition to the momentum transfer efficiency recorded from Earth, information about the DART crater ejecta plume will also be available and will be provided by the LICIACube \citep{Cheng2020}. LICIA (Light Italian Cubesat for Imaging of Asteroids) is the Italian Space Agency (ASI) contribution to the DART mission. The CubeSat will be carried by the DART spacecraft and will be released in the vicinity of the Didymos system before the impact. The main aim of the CubeSat is to take images of the ejecta plume, over a range of angle phases, at 136--163 s after the impact. 

The opacity of the crater ejecta plume as a function of height above the surface and time after the impact was shown to depend on target properties and can be used as a proxy for approximating the target strength \citep{Richardson2009, Holsapple2007, Cheng2020}. Current analytical models of the ejecta plume \citep{Holsapple2007, Cheng2020} assume that the ejected particles are launched from the impact point at a constant angle of 45$^\circ$ to the horizontal. However, this and previous work \citep{Luther2018} has shown that the ejection angle can vary by up to 30$^\circ$ depending on the target porosity or coefficient of internal friction and with azimuth in an oblique impact. For an accurate characterisation of the target properties, this variation in ejection ejection angle will need to be taken into account when comparing analytical models to impact ejecta plume observations. 

\section{Conclusions}

The DART mission will impact Didymos's satellite, Dimorphos, at an oblique angle, and deflect it by an amount detectable from Earth. The DART impact angle depends on the spacecraft's trajectory and the target slope at the point of impact, which is not known prior to the impact. The ejected material from oblique impacts is highly asymmetrical early in the cratering process, and this asymmetry influences the momentum transfer. iSALE-3D simulations of DART-like impacts at oblique angles show that the vertical momentum transfer efficiency (i.e., $\beta-1$) is similar for different impact angles. However, the imparted momentum is reduced as the impact angle decreases. For a 45$^\circ$ oblique DART-like impact, the momentum imparted to the target is expected to be reduced by up to 70\% compared to the momentum imparted from a vertical impact. Therefore, to achieve maximum deflection efficiency, it is desired that the DART spacecraft will hit Dimorphos at an angle as close to 90$^\circ$ as possible. 

For the cases we simulated here, the ejected momentum is not normal to the surface and there is a small downrange component to the net target momentum vector. However, the direction of the total momentum vector is observed to `straighten up' as crater growth becomes more symmetric at late times.

iSALE-2D simulations of vertical impacts show that the ejection angle of the crater ejecta is very sensitive to target properties, especially with target coefficient of internal friction, and can vary by up to 30$^\circ$. The ejection angle influences the geometry of the DART crater ejecta cone, which will be imaged by the LICIACube. 

The work presented here represents the first step towards an empirical scaling relationship for oblique impacts. This work can also be used as a framework to determine an analytical approximation of the vertical component of the ejecta momentum, $\beta-1$, given known target properties. While the derived model is in good agreement with laboratory experiments, iSALE-2D and iSALE-3D simulation results, further studies are needed to determine its limits of applicability. 

\section{Acknowledgements}

We gratefully acknowledge the developers of iSALE (www.isale-code.de), including Kai W\"unnemann, Dirk Elbeshausen, Boris Ivanov and Jay Melosh. This work has received funding from the UK's Science and Technology Facilities Council (STFC) (Grant ST/S000615/1) and from the European Union’s Horizon 2020 research and innovation programme, NEO-MAPP, under grant agreement No. 870377.

\section{Appendix}
Additional supporting information (tables, model outcomes) will be archived on GitHub and provided at the time of the publication as a DOI.
\newpage
\bibliography{refs.bib} 
\bibliographystyle{apa}  

\end{document}